\documentclass[a4paper,fleqn]{cas-dc}

\usepackage[authoryear,longnamesfirst]{natbib}
\usepackage[dvipsnames]{xcolor}
\usepackage{amsmath}
\usepackage{amssymb}
\usepackage{amsmath}
\usepackage{amsthm}
\usepackage{soul}
\usepackage{mathrsfs}
\usepackage{subfiles}
\usepackage[normalem]{ulem}
\usepackage{graphicx}
\usepackage[english]{babel}
\usepackage[T1]{fontenc}
\usepackage{verbatim}
\usepackage{ulem}
\usepackage{caption}
\usepackage{subcaption}
\usepackage{listings}
\usepackage[colorinlistoftodos]{todonotes}
\usepackage{cleveref}
\usepackage{lipsum}
\usepackage[percent]{overpic}
\usepackage{makecell}

\def\tsc#1{\csdef{#1}{\textsc{\lowercase{#1}}\xspace}}
\tsc{WGM}
\tsc{QE}

\newcommand{\bx}{\boldsymbol{x}}
\newcommand{\scrl}{\mathscr{L}}

\newcommand{\txbf}[1]{\textbf{#1}}

\usepackage{algorithm,algcompatible}

\algnewcommand\algorithmicto{\textbf{to}}
\algnewcommand\RETURN{\State \textbf{return} }

\newcommand{\INPUT}[1]{\STATE \textbf{Input}: #1 }
\newcommand{\OUTPUT}[1]{\STATE \textbf{Output}: #1 }

\newcommand{\ct}[1]{{\color{black}#1}}

\newcommand{\rwtwo}[1]{{\color{black}#1}}

\newcommand{\into}{\int_\Omega}
\newcommand{\defeq}{\overset{\text{\tiny def}}{=}}

\begin{document}
\let\WriteBookmarks\relax
\def\floatpagepagefraction{1}
\def\textpagefraction{.001}

\shorttitle{Inversion for Coefficient in Marine Lake Models}    

\shortauthors{A. Ho et~al.}  

\title [mode = title]{Adjoint-based Inversion for the Diffusion Coefficient in Marine Lake Models }  

\tnotemark[1,2] 

\tnotetext[1]{This work was partially supported by the National Science Foundation under Grant No. DMS-1840265.} 

\tnotetext[2]{We thank Gerda Ucharm, Sharon Patris, and Lori Colin at the Coral Reef Research Foundation, Palau, for generously providing data and methods from their long-term marine lake monitoring program in Palau that made this work possible.} 

%

\author[1]{Alex Ho}[orcid=0009-0000-2869-7637]

\cormark[1]


\ead{aho38@ucmerced.edu}


\credit{Conceptualization, Methodology, Software, Formal analysis, Investigation, Validation, Visualization, Writing – original draft, review \& editing}

\affiliation[1]{organization={Department of Applied Mathematics, University of California Merced},
            city={Merced},
            postcode={95343}, 
            state={CA},
            country={USA}}

\affiliation[2]{organization={Department of Quantitative and Systems Biology, University of California Merced},
            city={Merced},
            postcode={95343}, 
            state={CA},
            country={USA}}


\author[1]{Fran\c cois Blanchette}
\ead{fblanchette@ucmerced.edu}
\credit{Supervision, Conceptualization, Funding Acquisition, Methodology, Writing – review \& editing}

\author[1]{Noémi Petra}
\ead{npetra@ucmerced.edu}
\credit{Supervision, Conceptualization, Methodology, Writing – review \& editing}

\author[2]{Michael N Dawson}
\ead{mdawson@ucmerced.edu}
\credit{Resources, Supervision, Writing – review \& editing}






\cortext[1]{Corresponding author}



\begin{abstract}
    Marine lakes offer a unique opportunity to study how physical processes, such as turbulent mixing, biomixing, and tidal exchanges, regulate the vertical transport of heat, oxygen, and nutrients in relatively isolated ecosystems. The effective diffusion coefficient characterizes the overall mixing and is critical in models of marine lakes, but it is difficult to measure directly. This paper addresses the problem of inferring depth-dependent diffusion coefficient from synthetic and real measurements of vertical temperature and salinity profiles. To do so, an inverse problem governed by a marine lake model described by the screened-Poisson equation (similar to Helmholtz) is formulated. The inverse problem is formulated as a nonlinear least-squares optimization problem, where the cost functional quantifies the misfit between the observed and recovered profiles. A Tikhonov regularization term is added to \rwtwo{improve the conditioning of the discretized problem,} with a regularization parameter chosen using the L-curve method. 
    We solve this problem using an adjoint-based inexact Gauss-Newton method.
    To leverage the complementary information in these profiles, the formulation optimizes the fit across both datasets simultaneously.
    The accuracy of the reconstructed diffusion coefficient and the method's robustness to noise are investigated through comprehensive synthetic studies.
    Additionally, we contrast this framework with physics-informed neural networks, highlighting the advantages and limitations of each.
    When applied to real marine lake data, the adjoint method reconstructed a smaller diffusion coefficient in regions of strong density stratification, in line with physical expectations.
    The results indicate that the proposed approach provides a foundation for accurate modeling of marine lake dynamics.
\end{abstract}




\begin{keywords}
 Inverse problem \sep 
data-to-prediction \sep 
adjoint-based methods \sep
inexact Newton-CG method \sep
marine lake modeling 
\end{keywords}

\maketitle

\section{Introduction} \label{sec:intro}

\begin{figure}
        \centering
        \includegraphics[width=\linewidth]{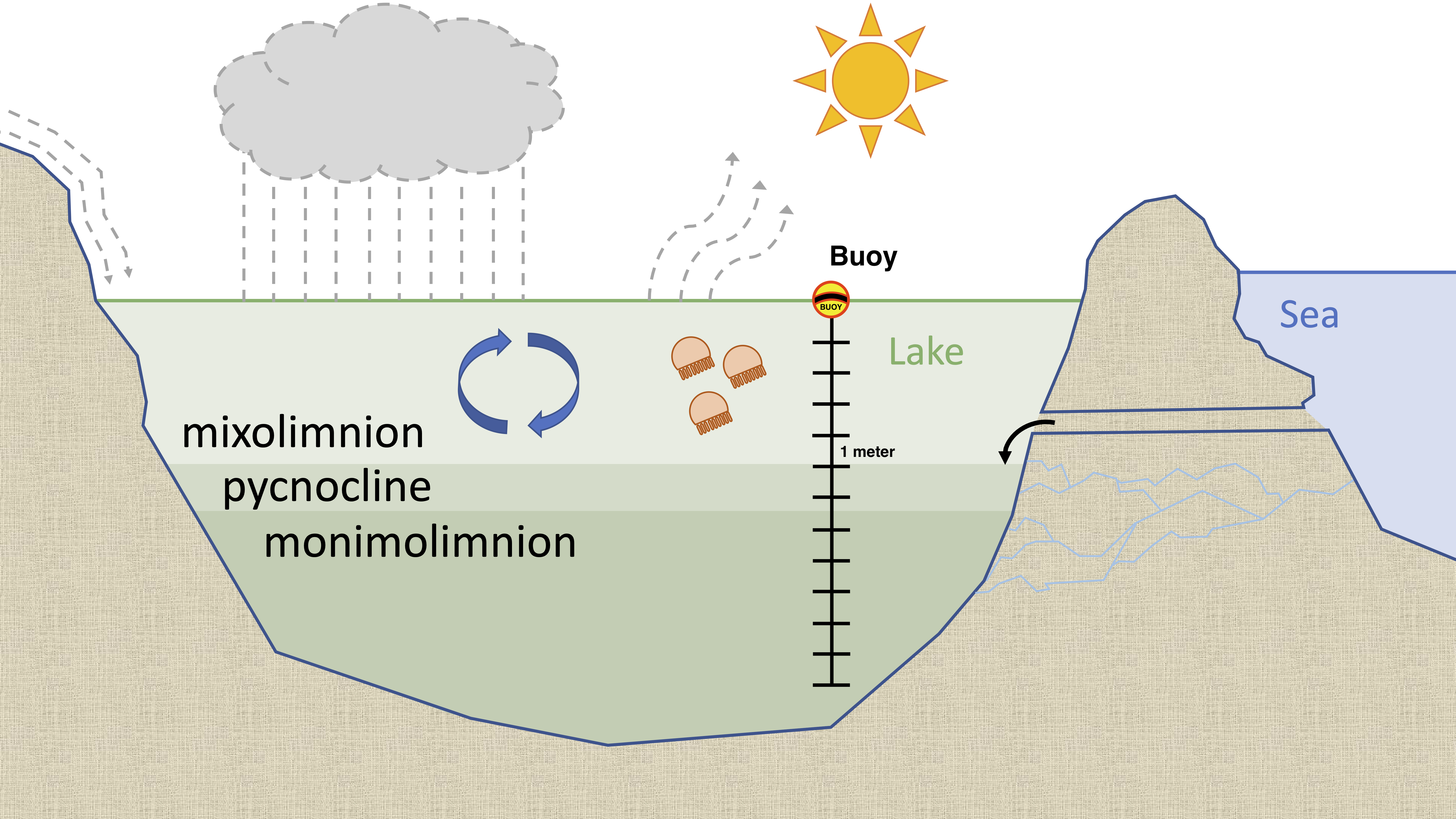}
        \caption{Schematic of a stratified marine lake. 
        Tidal exchange of brackish water occurs through subterranean tunnels and porous rock. Influx of freshwater from rain and runoff occurs at the lake's surface. 
        These processes contribute to the development of distinct density stratification, forming three layers: the mixolimnion, a generally well-mixed surface layer; the pycnocline, a narrow zone of steep density gradient that limits vertical exchange; the monimolimnion, a dense, often anoxic bottom layer. Measurements of temperature and salinity are taken with an array of data-loggers attached to a buoy, or a sonde deployed from the surface, at one-meter intervals. 
        }
        \label{fig:schematic_marine_lake}
    \end{figure}

Marine lakes are bodies of seawater with varying degrees of connection to the ocean via tunnels and porous rock \cite{hanzawa2012genetic,Dawson2005}. 
Lakes connected by shorter, wider, and more numerous tunnels tend to be well-mixed by tidal exchanges, whereas those with longer, narrower, and fewer tunnels are often vertically stratified. 
The isolation of these stratified lakes limits emigration and immigration, resulting in many endemic populations \cite{martin2006marine, dawson2001jellyfish}. 
Due to their relative physical and biological isolation from the open ocean, these stratified marine lakes may be considered ``natural aquaria,'' 
enabling detailed studies of how environmental changes affect marine ecosystems.
These effectively self-contained ecosystems are particularly well-suited for investigating how physical processes such as biomixing, tidal exchanges, and diffusion are affected by changing climatic conditions \cite{blanchette2020marine}.  
In particular, the degree of density stratification and its impact on mixing often play a crucial role in determining the physical and biological characteristics of these ecological systems \cite{strelkov2014marine}, with stratified and mixed lakes supporting distinct biotas \cite{rapacciuolo2019microbes}. 
Specifically, the vertical transport of quantities such as salinity, oxygen, and heat is dominated by turbulent motions and eddies that are hindered by density stratification, which often takes the form of a pycnocline (i.e., a region where the density changes abruptly, see \Cref{fig:schematic_marine_lake}).
The resulting mixing can be modeled by an effective diffusivity that is both determinant and difficult to quantify precisely \cite{houghton2018vertically,whalen2020internal}.

In stratified oceanic settings, mixing across a pycnocline 
is much weaker than in the mixed surface layer \cite{itoh2021vertical}; stratification strongly influences both the magnitude and vertical structure of the diffusivity \cite{whalen2012spatial}. 
In previous oceanic and lake studies, a constant effective diffusion coefficient above the pycnocline was often inferred from surface observations, turbulent measurements, and an understanding of internal wave processes \cite{garrett1979mixing,MACKINNON2013159}, and then adjusted to match observed density and velocity fields \cite{gregg1987diapycnal}.
Estimates have also been derived from turbulence experiments and Langmuir circulation under varying wind conditions \cite{filatov1981investigation, weller1988langmuir,thorpe1984effect}. 
In the context of marine lakes, 
time series of surface level 
plus depth profiles of
temperature and salinity were used to develop a numerical model to estimate the effective diffusion coefficient, with the coefficient held constant above and below the pycnocline and linearly varying within the pycnocline \cite{blanchette2020marine}.
Other approaches prescribe an empirical depth-dependent profile for diffusivity (typically a decaying exponential or Gaussian function) and determine its coefficients by fitting to local observational estimates \cite{kaufman1991depth,powell1974estimation,riley1988minlake}. 
Alternatively, the diffusivity may be estimated by directly integrating the conservation equations from observable data \cite{de2016estimation, li1973vertical}; however, this approach is highly sensitive to noise and requires smoothing to obtain a realistic estimate \cite{houghton2018vertically}.
While these methods provide useful insights, they are often ad hoc, 
limited in robustness, and sensitive to measurement noise, motivating the need for a more systematic approach.

In this paper, the depth-dependent effective diffusion coefficient is inferred from observed measurements of salinity and temperature profiles using a systematic and robust adjoint-based inverse method.
Inverse problems combine data with models to infer unknown parameters. They are used across a wide range of scientific and engineering applications, including, among others, image reconstruction \cite{Habring2022, Mardani2023AVP,deguchy2019image,ho2021convolution}, elasticity imaging \cite{Jadamba02122017}, and geophysical modeling, including glaciology \cite{Hartland2021,petra2014, petra2012} and seismology \cite{haber2014computational, haber2004inversion}.
The fundamental difficulty of solving inverse problems governed by 
differential equations (DEs) with infinite-dimensional parameters is that these problems are ill-posed, that is, their solution is not unique and is highly sensitive to observation errors
~\cite{Hadamard23,Ghattas_Willcox_2021}.
Indeed, in marine lake applications, the available data are sparse and noisy. 
In addition, there are limitations to which model parameters influence the data (i.e., low sensitivity in certain regions), which impacts the quality of the reconstruction (i.e., the ability to infer model parameters from the data).
To address these challenges, Tikhonov regularization \cite{vogelchap1,engl2015regularization} is incorporated in our adjoint-based inverse framework to avoid overfitting as we estimate the diffusion parameter iteratively.

Adjoint-based algorithms have been widely used to efficiently optimize inverse problems governed by differential equations, such as the scattering coefficient in the Helmholtz equation \cite{bui2012analysis, bui2012analysis2}, the wave speed coefficient in the acoustic wave equation \cite{bui2013computational}, the Lam\'e parameters in solid mechanics equations \cite{Jadamba02122017}, the basal boundary condition parameter in the nonlinear Stokes equations \cite{petra2014}, and the diffusion coefficient in the Poisson equation \cite{alexanderian2016fast, VillaPetraGhattas16}, to name a few. 
Here, the marine lake system under study is governed by a forced diffusion equation, and the focus is on the resulting steady state, described by the screened Poisson equation.
The gradient and the Hessian-apply required to efficiently solve the underlying optimization problem governed by the screened Poisson equation are derived using an adjoint-based technique that closely follows the formulation development for Poisson's equation of~\cite{VillaPetraGhattas16, petra2011}.
To the best of our knowledge, inversion for the spatially varying diffusion coefficient in the screened Poisson equation, or in the closely related Helmholtz equation, has not previously been reported.

\textbf{Contribution.} 
This paper has four key contributions. (1) The formulation of inverting a spatially-dependent effective diffusion coefficient from temperature and salinity profiles in marine lakes through
an adjoint-based algorithm is provided. This further refines
the work of  \cite{blanchette2020marine} and replaces spatially constant approximations obtained from observational data as in previous studies \cite{garrett1979mixing,MACKINNON2013159,gregg1987diapycnal,filatov1981investigation, weller1988langmuir,thorpe1984effect}.
(2) As an extension of the work of \cite{VillaPetraGhattas16, petra2011} on the Poisson equation, a method is obtained to efficiently invert for a spatially dependent diffusion coefficient in the screened Poisson equation, and its accuracy and robustness to noise are tested. This complements previous studies that invert for the scattering coefficient governed by the Helmholtz equation 
\cite{Crestel_2017,CHEN2021110114,haber2012, chaillat2012faims}.
(3) 
A systematic study and illustration of the benefits of incorporating multiple datasets using sensitivity and eigenvalue analyses are presented.
In addition, a diagnostic criterion is derived that indicates whether the model setup is sufficiently informative for successful inversion.
(4) The proposed method is applied to real observational data from a marine lake, demonstrating a successful inversion of a spatially varying effective diffusion coefficient and validating the robustness of the approach.

The remainder of this paper is organized as follows. \Cref{sec:gov_eq_num_setup} introduces the screened Poisson equation in the context of a marine lake and presents a corresponding 
derivation of the adjoint-based method, along with the sensitivity analysis derivation used in later experiments.
\Cref{sec:syn_prob} systematically investigates convergence and robustness and justifies incorporating more data through sensitivity analysis.
\Cref{sec:pinn_comparison} compares the adjoint-based inversion with a machine-learning approach, specifically physics-informed neural networks (PINNs).
\Cref{sec:marine_application} applies the setup to marine lake data, and the results of the inversion are presented and discussed. 
This includes developing a diagnostic tool for models.  
Section \ref{sec:conclusion} contains a summary of the reported findings and relevant conclusions.

\section{Governing equations and variational formulation} \label{sec:gov_eq_num_setup}
    \subsection{Forward problem} \label{sec:forward_problem_introduction}
    
    This section presents a mathematical model describing the distribution of physical properties, such as salinity and temperature, that affect the intrinsic dynamics of stratified marine lakes. 
    These systems are characterized by limited water exchange with the nearby sea through subterranean tunnels, so that, for example, they exhibit weak tides. 
    Limited communication with the sea makes marine lakes nearly enclosed marine ecosystems, akin to an island, that tend to host numerous endemic populations and species that are rare elsewhere \cite{dawson2015}. 
    Their detailed dynamics are complex \cite{blanchette2020marine} and comprise biological interactions \cite{Dawson2006,Dawson2005} as well as meteorological and fluid-dynamic factors \cite{martin2006marine}. 
    Here, the focus is on the last of the three: fluid dynamics.
    A partial differential equation that captures these dynamics is the forced diffusion equation
    \begin{equation}
        \begin{split}
        &\frac{\partial u}{\partial t}(\bx,t) - \nabla \cdot \left(D(\boldsymbol{x}) \, \nabla u(\boldsymbol{x},t)\right) \\
        &= \omega \, d(\boldsymbol{x},t)\, \left(u_{oce} - u(\boldsymbol{x},t)\right) + f(\boldsymbol{x},t),
        \end{split}
    \end{equation}
    where $u$ denotes temperature or salinity, $D(\bx)$ the effective diffusion coefficient, $\omega$ the exchange rate, $d(\bx)$ a distribution function, $u_{oce}$ the imposed ocean value, and $f(\bx, t)$ any additional sources or sinks. More details of these parameters will be provided in this section.
    
    Although time-dependent contributions can be important, the focus here is on a long, seasonal timescale where quasi-equilibrium is reached.
    As a result, a forced diffusion equation for the fluid property $u$ is studied. The steady-state formulation, namely a screened Poisson equation \cite{fetter2003theoretical} or an inhomogeneous Helmholtz-type equation, is considered:
    \begin{align}\label{eq:helm_main}
         - \nabla \cdot \left(D(\boldsymbol{x}) \, \nabla u(\boldsymbol{x})\right) = \omega \, d(\boldsymbol{x})\, \left(u_{oce} - u(\boldsymbol{x})\right) + f(\boldsymbol{x}).
    \end{align}
    In the present work, temperature or salinity are considered as quantities averaged horizontally over the lake's extent to reduce spatial dimensionality, so that the spatial variable $\bx$ can be thought of as the one-dimensional depth within the lake. 
    Nonetheless, to maintain general applicability, $\bx$ is treated as a vector in this derivation, and the differential operator $\nabla$ is used instead of $d/d\bx$. 
    This equation captures three dominant effects that govern the lake's dynamics. First, external forcing, $f(\bx)$, is imposed on the system and is independent of the value of the intrinsic property, $u$. When the quantity of interest is the water temperature, the forcing term can capture the long-term effect of solar irradiance imposed on the system. In the case where $u(\bx)$ represents salinity, the forcing term can account for runoff from rainfall. In general, such forcing is unevenly distributed with depth. It is assumed that $f(\bx)$ can be modeled independently and, therefore, is treated as given. 
    
    Second, exchanges with the nearby ocean are incorporated into the model. 
    These exchanges cause adjustments to the water property in a manner proportional to the difference between the ocean and lake property ($u_{oce} - u(\bx)$), where $u_{oce}$ is a characteristic value of the property described by $u$ that flows in from the ocean, assumed to be known and constant. 
    Although ocean-lake exchange occurs through porous rock and localized tunnels, its long-time effect is modeled as being distributed over the lake depth through a prescribed function $d(\bx)$, normalized to sum to one over the domain considered.
    Since a longer timescale is considered, $d(\bx)$ denotes the equilibrium distribution of exchanges over time.
    The magnitude of these exchanges can be assessed independently, in particular by considering tidal variations within the lake over time, and is quantified by the parameter $\omega$, with units of inverse time, which is assumed known. 
    Therefore, the exchange term takes the form $ \omega \, d(\boldsymbol{x})\, (u_{oce} - u(\boldsymbol{x}))$. 
    
    The third and most important factor included in the model is the diffusion of $u$ within the lake. 
    The strength of this diffusion is captured by the diffusion coefficient, $D(\bx)$, which is a determining factor in the dynamics of the lake \cite{blanchette2020marine}. 
    In practice, diffusion is driven primarily by turbulent, effectively random fluid motion, resulting in much more vigorous mixing than molecular motion.
    As a result, $D(\bx)$ is difficult to measure directly.
    Motivated by the difficulty of obtaining reliable measurements and the importance of inferring a spatially dependent diffusion coefficient, the present objective is to solve a DE-constrained optimization problem using adjoint-based techniques, thereby inferring $D(\bx)$ from temperature and salinity measurements.
    
    Despite the difficulty of measuring turbulent mixing, the diffusion coefficient is expected to exhibit certain properties.
    In a system where molecular diffusion is negligible, both temperature and salinity should diffuse identically since turbulent mixing affects both quantities in a similar manner.
    This mixing coefficient is generally depth-dependent, with stronger surface-layer fluid motion driven by wind and other meteorological factors \cite{botte2002model, riley1988minlake, filatov1981investigation}, resulting in larger values of $D$. 
    Turbulent mixing is also affected by variations in fluid density with depth. 
    A density-stratified water column, in which density increases with depth, will resist vertical mixing and yield smaller values of $D$ in regions where the density increases most rapidly.
    
    To ensure that it remains strictly positive, the effective diffusion coefficient is defined as  $D(\bx)~=~D_0e^{m(\bx)}$, where the prefactor $D_0$ carries the physical units and incorporates any prior knowledge of the expected magnitude of the diffusion coefficient. 
    The model is also non-dimensionalized with the following dimensionless variables:
    \begin{equation*}
        \bx^* = \frac{\bx}{L},\quad \nabla^* = L\nabla, \quad u^* = \frac{u - u_{oce}}{U}, \text{ and} \, f^*(\bx) = \frac{f(\bx)}{U},
    \end{equation*}
    where $L$ and $U$ are a length scale and the maximal variation of $u$, respectively. Additionally, even though quasi-equilibrium implies $\partial u/\partial t =0$, time units appear in the problem and must be rendered dimensionless. 
    To do so, a diffusive timescale $T = \frac{L^2}{D_0}$ is used as the time-scaling factor. 
    After some simplification, the non-dimensionalized version of the mathematical model is obtained 
    \begin{equation*}
        -\nabla^* \left(e^{m(L\bx^*)}\nabla^* u^*\right) = -\frac{\omega L^2}{D_0} \, d(L\bx^*) \, u^* + \frac{f(L\bx^*)L^2}{D_0}.
    \end{equation*}
    For ease of notation, let $m^*(\bx^*) = m(L\bx^*)$, $\omega^* = \frac{\omega L^2}{D_0}$, $d^*(\bx^*) = d(L\bx^*)$, and $f^*(\bx^*) = \frac{f(L\bx^*) L^2}{D_0}$.
    In what follows, all the asterisks are dropped to arrive at the following governing equation
    \begin{align}\label{eq:helm_nondim}
        -\nabla \cdot \left(e^{m(\bx)}\nabla u(\bx)\right) = -\omega \, d(\bx) \, u(\bx) + f(\bx)
    \end{align}
    subject to either Dirichlet-type boundary conditions or Neumann-type boundary conditions. 
    In the marine lake application, Dirichlet boundary conditions are chosen when $u$ represents the temperature and Neumann boundary conditions are chosen when $u$ represents the salinity. 
    Note that this non-dimensionalization results in a domain $\Omega = [0,1]$ and a normalized $u$.
    
    We thus consider a differential equation (DE) constrained optimization defined over the domain $\Omega$, with boundary conditions imposed on its boundary, $\partial \Omega$.
    The canonical approach is to first transform the DE constraint into its variational, or weak, form with
    test function denoted as $p(\boldsymbol{x})$.
    Here, $u(\bx)\in H^1(\Omega)$,
    and $p(\bx) \in H^1_0(\Omega)$ when $u$ satisfies a Dirichlet boundary condition; otherwise, $p(\bx) \in H^1(\Omega)$.
    To express the weak form, the $L^2$-inner product over the domain $\Omega$ and the boundary $\partial \Omega$ as $\langle \cdot, \cdot\rangle$ and $\langle \cdot,\cdot\rangle_{L^2(\partial \Omega)}$, are defined, respectively, as
    \begin{align*}
        \langle u, p\rangle = \into u\,p\,d\bx, \quad 
        {\langle v, p \rangle_{L^2(\partial \Omega)} = \int_{\partial \Omega} v\,p \,dS.}
    \end{align*}
    A key step in deriving the weak form is applying  Green's identity \cite{LoggMardalEtAl2012}, which allows 
    non-differentiability of 
    $\nabla u$ at a finite number of locations~\cite{Gockenbach06,zeidler2012applied,BeckerCareyOden81}. 
    Using these definitions, the weak form of \Cref{eq:helm_nondim} over the domain is given by 
    \begin{equation}\label{eq:weakform_DE_reduced_order}
    \begin{split}
        \langle e^{m}\,\nabla u , \nabla p \rangle   + \langle \omega \, d\,  u, p\rangle  - \langle f, p\rangle - \langle e^{m}\nabla u\cdot \textbf{n}, p\rangle_{L^2(\partial \Omega)} = 0, 
    \end{split}
    \end{equation}
    $\forall p \in V$, where $V = H_0^1(\Omega)$ or $H^1(\Omega)$.
    If a Dirichlet boundary condition is used, $p(\bx) \in H^1_0(\Omega)$, and the boundary integral term will vanish as $p=0$ on the boundary.
    Equation (\ref{eq:weakform_DE_reduced_order}) is the final form of the governing equation and will be used in the remainder of this article.
    Note {that in the one-dimensional case we have $\partial \Omega = \{x_L,x_R\}$. The boundary term {then} reduces to {evaluating} contributions at {the} two end points of the domain.}

\subsection{Inverse problem} \label{sec:single_data_VIP}
    In an inverse-problem framework, the objective is to infer an unknown parameter, in this case $m(\bx)$ after non-dimensionalization, that minimizes the mismatch between some given data $u_d(\bx)$ and the model prediction $u(\bx)$. 
    When solving inverse problems, it is advisable to use multiple datasets when available \cite{Crestel_2017}. 
    This paper will consider inverting for the same diffusion coefficient across multiple datasets and compare the quality of the inferred parameters with that obtained from a single dataset.
    A regularization term is included as a penalizing factor in the inverse solution to avoid overfitting the function $u(\bx)$ to the data. 
    Consider a general case where there are $N$ experimental data sets, each denoted with a subscript $i$, i.e., $u_{d,i}$, and denote the associated solution to the forward problem as $u_i$, with $i = 1,2,\hdots, N$. 
    The inverse problem governed by the forward problem~\eqref{eq:weakform_DE_reduced_order} can then be formulated as an optimization problem (in reduced space~\cite{borzi2011computational, antil2018frontiers}) as follows
    \begin{equation}\label{eq:cost_functional}
        \min_m\mathscr{J}(m) \defeq \sum_{i=1}^{N}\frac{\beta_i}{2}\langle u_i - u_{d,i},u_i - u_{d,i}\rangle + \frac{\gamma}{2}\langle \nabla m, \nabla m \rangle,
    \end{equation}
    where $u_i$ implicitly depends on the parameter $m$, and is the solution to the corresponding forward problem, namely
    \begin{equation}
        \langle e^{m} \nabla u_i, \nabla p\rangle  + \langle \omega \, d\,u_i,\,p\rangle - \langle e^{m}\nabla u_i\cdot \textbf{n}, p\rangle_{L^2(\partial \Omega)} = \langle f_i, p\rangle ,
    \end{equation}
    with $f_i$ and $p_i$ the forcing and test function of each problem, respectively.
    Note that each $u_i$ depends on $m$ implicitly through its governing equation and will satisfy its own boundary conditions. 
    The distribution $d(\bx)$ and the exchange rate $\omega$ are assumed to remain the same for all experiments, as mentioned in \cref{sec:forward_problem_introduction}.
    
    This paper only considers up to two experiments, so $N=1$ or $N=2$, and 
    the weighted coefficients are set to be equal, i.e. $\beta_i = \frac{1}{N}$, 
    for simplicity.
    However, the machinery works in general and can accommodate additional experiments as they become available.
    While other works have explored optimizing this coefficient to incorporate heterogeneous data or reduce computational costs \cite{Crestel_2017, haber2012,Haber2014}, this is beyond the scope of the current study.
    Finally, the regularization parameter $\gamma$  weighs the contribution of the regularization (in this paper, we use Tikhonov regularization). 
    Here, the selection of $\gamma$ is made using 
    the L-curve method, which does not require any prior knowledge of the data or the solution of the inverse problem \cite{hansen2001lcurve,vogelchap1}.
    
    \subsection{Inexact Newton-CG applied to solve the inverse problem} \label{sec:newt_inv}
    To ensure that the DE constraint is satisfied, a Lagrangian functional is constructed by combining the cost functional from (\ref{eq:cost_functional}) with the variational form of the DE from \Cref{eq:weakform_DE_reduced_order}, here shown for $N=1$ for simplicity:
    \begin{equation}\label{eq:lagrangian_general}
    \begin{split}
        \mathscr{L}(u,m,p) &\defeq {\frac{1}{2} \langle u - u_d,u - u_d\rangle  + \frac{\gamma}{2}\langle \nabla m, \nabla m\rangle} \\
        &+ \langle e^{m}\,\nabla u , \nabla p \rangle   + \langle \omega \, d\,  u, p\rangle  - \langle f, p\rangle \\ &- \langle e^{m}\nabla u\cdot \textbf{n}, p\rangle_{L^2(\partial \Omega)} ,
    \end{split}
    \end{equation}
    where $p(\bx)$ is the Lagrange multiplier. As stated in~\cite{petra2011, Neitzel2009}, the optimality condition requires that the variation of the Lagrangian with respect to each input function, evaluated at the optimal solution, vanishes; that is, 
    \begin{equation}
        \label{eq:state_eq_general}
        \begin{split}
    	\mathscr{L}_p(u,m,p)(\tilde{p}) = &\langle e^{m} \, \nabla u ,\nabla\tilde{p}\rangle + \langle\omega \, d\, u , \tilde{p} \rangle - \langle f,\tilde{p}\rangle \\ &- \langle e^{m}\nabla u\cdot \textbf{n}, \tilde{p}\rangle_{L^2(\partial \Omega)}  = 0, 
        \end{split}
    \end{equation}
    \begin{equation}
        \label{eq:adj_eq_general}
        \begin{split}
    	\mathscr{L}_u(u,m,p)(\tilde{u}) = &\langle u - u_d, \tilde{u} \rangle + \langle e^{m}\nabla p , \nabla \tilde{u}\rangle \\&+ \langle \omega \,d\,p,\tilde{u}\rangle= 0, 
        \end{split}
    \end{equation}
    \begin{equation}
        \label{eq:grad_eq_general}
        \begin{split}
        \mathscr{L}_m(u,m,p)(\tilde{m}) &= \gamma\langle\nabla m, \nabla \tilde{m}\rangle + \langle e^{m}\tilde{m} \, \nabla u , \nabla p\rangle \\ &=  0, 
        \end{split}
    \end{equation}
    where the 
    variation of the Lagrangian is taken with respect to the given subscript, and $\tilde{u}$, $\tilde{m}$, and $\tilde{p}$ are the variations of $u$, $m$, and $p$, respectively. The three formulae given above are the \textit{state}, \textit{adjoint}, and \textit{gradient} equations, respectively.
    Note that the boundary integral does not appear in the adjoint equation nor the gradient, since the term vanishes if: (1) A Dirichlet boundary condition is given and $p\in H^1_0(\Omega)$ vanishes at the boundary, or (2) a Neumann boundary condition is given, i.e. $h=e^m\nabla u \cdot \textbf{n}$ is fixed, and $\langle h, p\rangle_{L^2(\partial \Omega)}$ does not depend on $u$ or $m$, thus the variation is $0$.
    Also note that the state equation only has a boundary integral if a Neumann boundary condition is used; otherwise, the term will vanish because $p$ vanishes on the boundary.
    
    The optimization problem is solved using an inexact Newton conjugate gradient (INCG) method \cite{Ghattas_Willcox_2021}, and hence in what follows, we show the derivation of the Hessian (action to a vector)~\cite{borzi2011computational,NoceWrig06}.
    To obtain an expression for the Hessian, a meta-Lagrangian functional  \cite{Ghattas_Willcox_2021} that consists of the state equation, the adjoint equation, and the gradient is constructed:
    \begin{equation}
        \label{eq:metafunc}
        \begin{split}
        \mathscr{L}^H(u,m,p)(\hat{u},\hat{m},\hat{p}) \defeq &\scrl_p(u,m)(\hat{p}) + \scrl_u(u,m,p)(\hat{u}) \\ &+ \scrl_m(u,m,p)(\hat{m}),
        \end{split}
    \end{equation}
    where $\hat{u}$, $\hat{m}$, and $\hat{p}$ are the Lagrange multipliers. Note that the hat functions were previously indicated as $\tilde{u}$, $\tilde{m}$, and $\tilde{p}$ as the variations of $u$, $m$, and $p$, respectively. 
    The change in notation represents the transition from test function to Lagrange multiplier when state, adjoint, and gradient are enforced in Lagrangian $\mathscr{L}^H$ \cite{Ghattas_Willcox_2021}.
    Similar to the first variations 
    (from (\ref{eq:state_eq_general})-(\ref{eq:grad_eq_general})), Hessian information
    can be found by taking the variation of (\ref{eq:metafunc}) with respect to $m$, $u$, and $p$, namely
    \begin{equation}\label{eq:inc_adj_eq_general}
        \begin{split}
        \mathscr{L}_u^H(\hat{m}, \hat{u},\hat{p}) = &\langle\hat{u},\tilde{u}\rangle + \langle  e^m \hat{m}\nabla p, \nabla \tilde{u} \rangle + \langle e^m \nabla \hat{p}, \nabla \tilde{u}\rangle \\&+ \langle {\omega} \, d \, \hat{p} , \tilde{u}\rangle , \\
        \end{split}
    \end{equation}
    \begin{equation}\label{eq:hessian_eq_general}
        \begin{split}
        \mathscr{L}_m^H(\hat{m}, \hat{u}, \hat{p}) = &\langle e^m\tilde{m} \nabla p , \nabla \hat{u}\rangle + \langle e^m\tilde{m} \hat{m}\nabla u , \nabla p\rangle \\&+ {\gamma}\langle\nabla \hat{m} , \nabla \tilde{m} \rangle +  \langle e^m\tilde{m}\nabla u , \nabla \hat{p}\rangle, \\
        \end{split}
    \end{equation}
    \begin{equation}\label{eq:inc_state_eq_general} 
        \begin{split}   
        \mathscr{L}_p^H(\hat{m},\hat{u},\hat{p}) =   &\langle e^m\nabla \hat{u}, \nabla \tilde{p} \rangle + \langle {\omega} \, d\, \hat{u},\tilde{p}\rangle \\&+ \langle e^m \hat{m}\nabla u , \nabla \tilde{p}\rangle , 
        \end{split}
    \end{equation}
    which are known as \textit{incremental-adjoint}, 
    \textit{Hessian-apply}, and \textit{incremental-state}, 
    respectively~\cite{Ghattas_Willcox_2021}.

\begin{algorithm}[t]
        \small
        \caption{Inexact Newton-CG applied to solve the optimization problem~\eqref{eq:cost_functional} with $N=1$.}
        \label{alg:method_alg}
        \begin{algorithmic}[1]
        \INPUT {Initial guess $\textbf{m}_0$, }
        \OUTPUT {\textbf{m}, \textbf{u}}
        \STATE $j \leftarrow0$
        \WHILE{\textbf{{not converged}}}
        \STATE $\textbf{u}_j \leftarrow \left(\textbf{A} + \omega\hat{\textbf{\ct{M}}}\right)\textbf{u}_j = \textbf{f}$
        \STATE $\textbf{p}_j \leftarrow \left(\textbf{A}^\intercal + \omega\hat{\textbf{\ct{M}}}^\intercal\right)\textbf{p}_j = -\textbf{M}\left(\textbf{u}_j - \textbf{u}_d\right)$
        \STATE $\textbf{g}_j = {\textbf{R}\textbf{m}_j} + \textbf{C}(\textbf{p}_j)\textbf{u}_j$ 
        \STATE $\text{tol}_{cg} = \min(0.5, \lVert \textbf{g}_j\rVert\ct{/\lVert \textbf{g}_0}\rVert)$
        \STATE $\hat{\textbf{m}}_0 = \textbf{0}$
        \STATE $i\leftarrow 0, \hat{\textbf{m}} \leftarrow 0$
        \WHILE{$\lVert\textbf{H}\hat{\textbf{m}}_i + \textbf{g}_j\rVert > \text{tol}_{cg}$}
        \STATE $\hat{\textbf{u}}_i \leftarrow (\textbf{A} + \ct{\omega}\hat{\textbf{M}})\hat{\textbf{u}}_i = \hat{\textbf{C}}^\intercal\hat{\textbf{m}}_i$
        \STATE $\hat{\textbf{p}_i} \leftarrow (\textbf{A}^\intercal + \ct{\omega}\hat{\textbf{M}}^\intercal)\hat{\textbf{p}}_i = \left(\textbf{M}\hat{\textbf{u}}_i + \textbf{C}^\intercal\hat{\textbf{m}}_i\right)$
        \STATE $\hat{\textbf{m}}_{i+1} \leftarrow \textbf{H}\hat{\textbf{m}}_i = -\textbf{g}_j$ 
        \STATE $i \leftarrow i+1$ 
        \ENDWHILE
        \STATE $\alpha_j \leftarrow \text{Armijo}(\textbf{m}_j, \hat{\textbf{m}}_i)$ 
        \STATE ${\textbf{m}}_{j+1} \leftarrow {\textbf{m}}_j + \alpha_j\hat{\textbf{m}}_i$ 
        \STATE $j \leftarrow j+1$ 
        \ENDWHILE
        
        \end{algorithmic}
    \end{algorithm}
\subsection{Discretization and algorithm} \label{sec:fem_disc}

The forward problem, \Cref{eq:helm_nondim}, is solved using the Galerkin finite element method (FEM) \ct{with $N_x$ elements} to approximate functions and their integrals.
In what follows, we consider the finite-dimensional subspace spanned by continuous Lagrange nodal basis functions $\phi_i$ with \ct{$N_x+1$ denoting the number of nodal basis functions}, i.e., $V = \text{span}(\phi_1(\bx),\hdots,\phi_{\ct{N_x+1}}(\bx)) \subset L^2(\Omega)$. 
The solution $u$ and test function $p$ are approximated with the same basis functions, e.g., $u(\bx) = \sum_{i=1}^{\ct{N_x+1}}u_i\phi_i(\bx)$ and $p(\bx) = \sum_{i=1}^{\ct{N_x+1}}p_i\phi_i(\bx).$
Then, letting \textbf{u} and \textbf{p} represent the vectors of coefficients of the basis functions, i.e., $\textbf{u}=(u_1,\hdots, u_{\ct{N_x+1}})$ and $\textbf{p} = (p_1,\hdots,p_{\ct{N_x+1}})$, \Cref{eq:state_eq_general} can be written as the linear system:
\begin{equation} \label{eq:linear_weakform}
\begin{split}
    \Tilde{\textbf{p}}^\intercal \textbf{A}\textbf{u} + \omega \Tilde{\textbf{p}}^\intercal \hat{\textbf{M}}\textbf{u} - \Tilde{\textbf{p}}^\intercal \textbf{f} = \Tilde{\textbf{p}}^\intercal \left(\textbf{A}\textbf{u} + \omega \hat{\textbf{M}}\textbf{u} - \textbf{f} \right) = 0,
\end{split}
\end{equation}
where the stiffness matrix \textbf{A}, the mass matrix $\hat{\textbf{M}}$, and  the forcing vector \textbf{f} are defined as 
\begin{align*}
    &\textbf{A}_{ik} =\langle e^{m(\bx)}\nabla \phi_i(\bx), \nabla \phi_k(\bx)\rangle ,  \\ &\hat{\textbf{M}}_{ik} = \langle d(\bx)\phi_i(\bx),\phi_k(\bx)\rangle, \\ &\textbf{f}_k = \langle f(\bx),\phi_k(\bx)\rangle,
\end{align*}
where $i,k=1,\hdots,\ct{N_x+1}$. 
\ct{Since $m(\bx)$ and $d(\bx)$ are also represented in the finite-element basis, the element integrals {for $\textbf{A}$ and $\hat{\textbf{M}}$ }
are evaluated during matrix assembly using standard numerical quadrature in FEniCS \cite{logg2012}.} {We also note that both of these matrices are symmetric.}
For demonstration, \textbf{u} is assumed to satisfy the Dirichlet boundary condition so that the boundary term will disappear due to $p(\bx)\in H^1_0(\Omega)$. 
The linear system given in \Cref{eq:linear_weakform} needs to hold for all $\tilde{\txbf{p}}$, and, therefore,
\begin{align}
    \ct{\textbf{g}_p := (\textbf{A} + \omega \hat{\textbf{M}})\textbf{u} -  \textbf{f}=0. }\label{eq:state_eq_disc}
\end{align}
By applying a similar process to the adjoint equation and the gradient, the linear systems of equations for \Cref{eq:adj_eq_general} and the left-hand-side of \Cref{eq:grad_eq_general} become
\begin{align}
    \textbf{g}_u &:=(\textbf{A}^\intercal + \omega {\hat{\textbf{M}}}^\intercal) \textbf{p} + \textbf{M}(\textbf{u} - \textbf{u}_d) =0,\label{eq:adj_eq_disc}\\
    \textbf{g}_m &:= \textbf{R}\textbf{m} +\textbf{C}\textbf{u},\label{eq:grad_eq_disc}
\end{align}
where the mass matrix \textbf{M}, the stiffness matrix \textbf{R}, and the matrix \textbf{C} are defined as
\begin{align*}
    &\textbf{M}_{ik} = \langle \phi_i(\bx),\phi_k(\bx)\rangle, \\ &\textbf{R}_{ik} = \langle \nabla \psi_i(\bx), \nabla \psi_k(\bx)\rangle,\\ &\textbf{C}_{ik} = \langle e^{m(\bx)}\psi_i(\bx)\nabla p(\bx),\nabla \phi_k(\bx)\rangle,
\end{align*}
where $i,k=1,\hdots,\ct{N_x+1}$,
\ct{and $\psi_i(\bx)$ denotes the finite-element basis used to represent the parameter $m(\bx)$. In this work, the same basis functions are used to represent the parameter and the state variable, i.e., $\phi(\bx) = \psi(\bx)$, but the notation is kept distinct for generality. Note that the entries of \txbf{C} depend on $p(\bx)$.
{
{Equations (\ref{eq:state_eq_disc}), (\ref{eq:adj_eq_disc}), and (\ref{eq:grad_eq_disc})}  define the discrete state, adjoint, and gradient residuals, respectively.
At a stationary point, all three residuals vanish.
}
}

At each optimization iteration, given a parameter $\textbf{m}$, we solve the forward problem~\eqref{eq:state_eq_disc} for $\textbf{u}$, and then given $\textbf{m}$ and $\textbf{u}$ we solve the adjoint problem~\eqref{eq:adj_eq_disc} for the adjoint \textbf{p}.
Once these equations are solved, the gradient shown in (\ref{eq:grad_eq_disc}) is evaluated, which is generally non-zero unless the optimality condition is satisfied.
Since Newton's method is used, Hessian information is required to determine the \ct{Newton} search direction. The same discretization process \ct{is applied to}
\eqref{eq:inc_adj_eq_general}--\eqref{eq:inc_state_eq_general}, yielding the discretized incremental-adjoint equation, Hessian-apply, and incremental-state equation. \ct{If we define $\textbf{K} = (\bf{A}+\omega\hat{\bf{M}})$, also a symmetric matrix,} these can be organized as 
\begin{align}\label{eq:second_var_linear_sys}
    \left[
    \begin{array}{ccc}
        \textbf{M} & \textbf{C}^\intercal & \ct{\textbf{K}^\intercal}\\
        \textbf{C} & (\textbf{D} + \textbf{R}) & \hat{\textbf{C}} \\
        \ct{\textbf{K}} & \hat{\textbf{C}}^\intercal & \textbf{0}
    \end{array}
    \right]
    \left[
    \begin{array}{c}
        \hat{\textbf{u}}\\
        \hat{\textbf{m}}\\
        \hat{\textbf{p}}
    \end{array}
    \right] = -
    \left[
    \begin{array}{c}
         \mathbf{g}_u \\  \textbf{g}_m\\ \textbf{g}_p
    \end{array}
    \right],
\end{align}
where $\mathbf{g}_u$, $\mathbf{g}_m$, $\mathbf{g}_p$ 
\ct{denote the left-hand-side of \cref{eq:adj_eq_disc}, \cref{eq:grad_eq_disc}, and \cref{eq:state_eq_disc}, respectively.}
The matrices \textbf{D} and $\hat{\textbf{C}}$ are defined as
\begin{align*}
    &\textbf{D}_{ik} = \langle e^{m(\bx)}\psi_i(\bx)\psi_k(\bx)\nabla u(\bx),\nabla p(\bx)\rangle, \\ &\hat{\textbf{C}}_{ik} = \langle e^{m(\bx)}\psi_i(\bx)\nabla u(\bx),\nabla \phi_k(\bx)\rangle, 
\end{align*}
where $i,k=1,\hdots,\ct{N_x+1}$. Here, we work in a reduced space framework~\cite{borzi2011computational}, i.e., we assume that $\textbf{u}$ and $\textbf{p}$ satisfy the state and the adjoint equations such that $\textbf{g}_u = \textbf{g}_p = 0$. 
Denoted by \textbf{H}, the reduced Hessian applied to a direction $\hat{\txbf{m}}$ is given by
\begin{align}\label{eq:hessian_disc}
    \textbf{H}\hat{\textbf{m}} = \textbf{C}\hat{\textbf{u}} + (\textbf{D} + \textbf{R})\hat{\textbf{m}} + \hat{\textbf{C}}\hat{\textbf{p}},
\end{align}
where the incremental state and adjoint variables, $\hat{\textbf{u}}$ and $\hat{\textbf{p}}$, are obtained from \ct{solving the first {and third} block rows} 
of~\eqref{eq:second_var_linear_sys}, namely by solving the incremental state and  adjoint equations, respectively, i.e., 
\begin{align}
    \hat{\textbf{p}} &= \ct{\textbf{K}^{-\intercal}}\left(\textbf{M}\hat{\textbf{u}} + \textbf{C}^\intercal\hat{\textbf{m}}\right),\label{eq:inc_adj_eq_disc} \\
    \hat{\textbf{u}} &= \ct{\textbf{K}^{-1}}\hat{\textbf{C}}^\intercal\hat{\textbf{m}}.\label{eq:inc_state_eq_disc}
\end{align}
This results in the following reduced linear system for the Newton \ct{search direction} $\hat{\textbf{m}}$
\begin{align}\label{eq:newt_linear_system}
    \textbf{H}\hat{\textbf{m}} = -\textbf{g}_m.
\end{align}
To expose the
structure of the Hessian operator, below we show the explicit form of the Hessian matrix (though this is never formed in practice)
\begin{equation}\label{eq:reduced_hessian}
\begin{split}
    \textbf{H} := &\textbf{C}\ct{\bf{K}^{-1}}\hat{\textbf{C}}^\intercal + (\textbf{D} + \textbf{R}) + \hat{\textbf{C}}\ct{\bf{K}^{-\intercal}}\textbf{M}\ct{\bf{K}^{-1}}\hat{\textbf{C}}^\intercal \\&+ \hat{\textbf{C}}\ct{\bf{K}}^{-\intercal}\textbf{C}^\intercal.
\end{split}
\end{equation}
\ct{
The terms $\textbf{C}{\bf{K}^{-1}}\hat{\textbf{C}}^\intercal$ and $\hat{\textbf{C}}{\bf{K}}^{-\intercal}\textbf{C}^\intercal$ form a transpose pair that arises from {differentiation,} 
ensuring that the sum of the pair results in a symmetric operator. 
}

The reduced-Hessian is composed of two terms. The first is the so-called Hessian of the misfit term (incorporating information about the data fitting), and the second is the Hessian corresponding to the regularization term, \ct{here given by \txbf{R}}. 
In mathematical notation, this reads $\textbf{H} =\textbf{H}_{\text{misfit}} + \textbf{R}$,
\ct{where 
$$\textbf{H}_{\text{misfit}} =\textbf{C}\ct{\bf{K}^{-1}}\hat{\textbf{C}}^\intercal + \textbf{D} + \hat{\textbf{C}}\ct{\bf{K}^{-\intercal}}\textbf{M}\ct{\bf{K}^{-1}}\hat{\textbf{C}}^\intercal + \hat{\textbf{C}}\ct{\bf{K}}^{-\intercal}\textbf{C}^\intercal.$$
}
Equivalently, we can write this as $\textbf{H} = \ct{\textbf{R}}\left(\textbf{H}_p  + \textbf{I}\right)$, where  
\begin{equation}
\label{eq:precond_hessian}
\textbf{H}_p = \ct{\textbf{R}}^{-1}\textbf{H}_{\text{misfit}}
\end{equation}
is the so-called regularization preconditioned Hessian of the misfit term.

In this paper, we use the conjugate gradient (CG) method to solve the Newton system iteratively, using \ct{$\textbf{R}$} as a preconditioner. The procedure proceeds as follows.
\rwtwo{Starting with an initial guess for the discretized parameter~$\textbf{m}$, Newton's method iteratively updates this parameter by 
$\textbf{m}_{\text{j+1}} = \textbf{m}_{\text{j}} + \alpha \hat{\textbf{m}}$,
where $\textbf{m}_{\text{j}}$ is the current parameter, and the Newton direction $\hat{\textbf{m}}$ is approximated by solving the linear system~\eqref{eq:newt_linear_system} via CG. 
To reduce the number of linearized DE solves per CG iteration, we use an inexact Newton-CG method that terminates the CG iterations when the norm of the residual of the linear system~\eqref{eq:newt_linear_system} drops below a tolerance that is proportional to the norm of the gradient, i.e., we use a tolerance defined by~\cite{VillaPetraGhattas16,NoceWrig06,kelley1999iterative}  }
\begin{align}
    \text{tol}_{cg} = \min\left(0.5, \frac{\lVert\textbf{g}_j\rVert}{\lVert\textbf{g}_0\rVert}\right).
    \label{eq:tolcg}
\end{align} 
To improve the robustness of the Newton iterations, we use a backtracking Armijo line-search~\cite{NoceWrig06} to find an appropriate step length $\alpha$. This step length ensures that the cost functional decreases sufficiently at each iteration. The Newton iterations are terminated when the norm of the gradient falls below a threshold. We note that
Gauss-Newton approximation of the reduced-Hessian is employed for the first five iterations to help with convergence. 
{This Gauss-Newton approximation of the Hessian is obtained by ignoring the matrix blocks \textbf{C} and \textbf{D}~\cite{Ghattas_Willcox_2021}. 
{This approximation is appropriate in the small-noise regime, since the data misfit drives the adjoint equation.
We note that a small data misfit leads to an adjoint that is small in magnitude, and therefore the Hessian terms involving \txbf{C} and \txbf{D} become less important.}}
Readers interested in more background on the Gauss–Newton approximation and related second-order methods may consult, e.g., {\cite{NoceWrig06, kelley1999iterative} or 
in the context of inverse problems governed by PDEs 
~\cite{Ghattas_Willcox_2021, petra2011}.}
Finally, we note that for the range of values we tried, the exact choice of the initial condition did not affect the recovered parameter.

\subsection{Sensitivity analysis} \label{sec:sens_analysis}
To better understand the quality of the inverse problem solution, this section presents a sensitivity analysis in discrete form. 
The sensitivity of the solution $u$ to the parameter of interest, $m$, informs how well the parameter can be reconstructed from data $u_d$. 
In a least-squares setting, this quantifies the amount of information that can be extracted from the data.
The unknown $m(\bx)$ is approximated as 
\begin{align}\label{eq:disc_m_normalized}
    m(\bx) = \sum_{i=1}^{\ct{N_x+1}} m_i \psi_i(\bx) ,
\end{align}
where $\psi_i(\bx)$ is a finite-element basis function.
From the linear system shown in \Cref{eq:state_eq_disc}, the discretized residual is defined as
\begin{equation}
    \textbf{r} \defeq \textbf{A}\textbf{u} + \omega \hat{\textbf{M}}\textbf{u} - \textbf{f} =  \textbf{r}(\bf{\mu},\textbf{u}) = 
    \ct{\textbf{A}(\textbf{m})} \textbf{u} +\omega\hat{\textbf{M}}\textbf{u} - \textbf{f},
    \label{eq:def_r}
\end{equation}
where \textbf{A} is a matrix that depends on \textbf{m} as shown in the previous subsection. \ct{To clarify the notation in the upcoming differentiation process, a temporary variable is introduced, $\bf{\mu} = \textbf{m}$, and we replace $\textbf{A}(\textbf{m})$ with $\textbf{A}(\bf{\mu})$ in the previous equation.} Since each component of the vector \textbf{u} depends implicitly on each component of vector \textbf{m}, differentiating the $k^{th}$ residual value with respect to $m_j$ gives:
\begin{equation}
\begin{split}
    \frac{\partial {r}_k}{\partial {m}_j} =  &\sum_{i=1}^{\ct{N_x+1}} \frac{\partial {r}_k}{\partial {u}_i} \frac{\partial {u}_i}{\partial {m}_j}  + \sum_{i=1}^{\ct{N_x+1}} \frac{\partial {r}_k}{\partial \mu_i}\frac{\partial \mu_i}{\partial {m}_j} \\&=  \sum_{i=1}^{\ct{N_x+1}} \frac{\partial {r}_k}{\partial {u}_i} \frac{\partial {u}_i}{\partial {m}_j}  + \frac{\partial {r}_k}{\partial \mu_j}  = 0,   
\end{split}
\end{equation}
where $j,k = 1,2,\hdots, \ct{N_x+1}.$
Note that since $\mu = \textbf{m}$, ${\partial \mu_i}/{\partial {m}_j} = 1$ when $i=j$, and ${\partial \mu_i}/{\partial {m}_j} = 0$ when $i\neq j$, 
which reduces the sum to only one component, namely $\frac{\partial \textbf{r}}{\partial \mu_j}$. 
If \textbf{u} and \textbf{m} used for evaluating the residual satisfy the weak form of the DE, the residual equates to zero. Then the sensitivity vector of $\frac{\partial \textbf{r}}{\partial {m}_j}$ can be written as
\begin{align}
    \frac{\partial \bf{r}}{\partial {m}_j} = \left(\bf{A} + \omega \hat{\bf{M}}\right) \frac{\partial \textbf{u}}{\partial {m}_j} + \frac{\partial \bf{r}}{\partial \mu_j} = 0.
\end{align}
Therefore, the sensitivity equations read:
\begin{align}\label{eq:sen_eq_cont}
    \left(\bf{A} + \omega \hat{\bf{\ct{M}}}\right)\frac{\partial \textbf{u}}{\partial m_j} = -\frac{\partial \bf{r}}{\partial \mu_j}= -  \hat{\textbf{C}}_j,  
\end{align}
where $ j = 1, 2, \dots , \ct{N_x+1},$ and $\hat{\textbf{C}}_j$ is the $j^{th}$ row of $\hat{\textbf{C}}$ matrix defined in the previous section. 
By finding ${\partial\textbf{u}}/{\partial m_j}$ for all $j$, a sensitivity matrix, $\tfrac{\partial \textbf{u}}{\partial \textbf{m}}$, can be constructed, 
where each entry represents the sensitivity of ${u}_i$ with respect to ${m}_j$.
An example of this is shown in \Cref{fig:flat_u_effect_plot} of \Cref{sec:sens_for_single_n_mult}.
Please refer to \cite{ho2024inv} for a more complete derivation.

\section{Results for model test problems} \label{sec:syn_prob}

    \subsection{Model problems} \label{sec:num_setup}
    \begin{figure*}
        \centering
        \vspace{0.2cm}
        \begin{overpic}[width=1\linewidth]{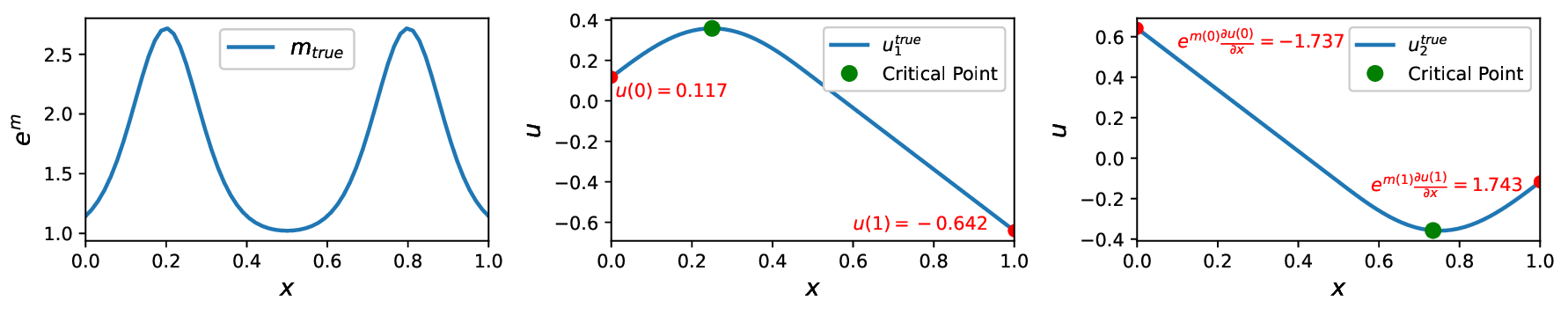}
            \put(7.5,19.5){\footnotesize A.}
            \put(41,19.5) {\footnotesize B.}
            \put(75,19.5) {\footnotesize C.}
        \end{overpic}
        \caption{ 
        The true parameter field $m_{true}$ (Panel A) and the state solutions $u^{true}_i$ defined in \Cref{eq:synth_true_sol} that satisfy Dirichlet (Panel B) and Neumann (Panel C) boundary conditions. Each state solution has a critical point in a different part of the domain, as shown by the green dots.
        }
        \label{fig:synthetic_true_solution}
    \end{figure*}
    To evaluate the performance of the proposed inversion framework,  this section presents two one-dimensional synthetic model problems that mimic the physical characteristics of temperature and salinity profiles. 
    Specifically, each model describes the steady-state quantity $u_i(x)$ that satisfies the Dirichlet ($i=1$, mimicking temperature) or Neumann ($i=2$, mimicking salinity) boundary conditions.
    These models serve as test cases for convergence analysis with mesh refinement, robustness testing against noise, and comparison between different optimization strategies.

    \textbf{Forward problem setup.} 
    Here, the normalized form of the model problems is defined on the domain $\Omega \in  [0,1]$.
    The governing equations of the forward problems are
    \begin{align}
        -\nabla \cdot ( e^{m(x)} \, \nabla u_i(x)) + \omega \, d(x)\,  u_i(x)\,  = f_i(x),\label{eq:reminder_us_gov_eq}
    \end{align}
    for $i=1$ or $i=2$.
    
    \textbf{Synthetic measurements.}
    To simulate observational uncertainty, Gaussian noise is added to the true solutions: 
    \begin{align} \label{eq:synth_data}
        \textbf{u}_d = \mathcal{F}(m) + \varepsilon, \quad \text{where }\textbf{u}_d = \begin{bmatrix}
            u_{d1} \\ u_{d2}
        \end{bmatrix}, \quad \varepsilon = \begin{bmatrix}
            \varepsilon_1 \\ \varepsilon_2
        \end{bmatrix},
    \end{align}
    and $\mathcal{F}(m)$ \ct{denotes the} parameter-to-observation map, 
    \ct{i.e., 
    $\mathcal{F}(m)$ outputs a vector of sampled observations whose components are $u_{1}^{true}(x_j)$ and $u_{2}^{true}(x_j)$ evaluated at each discretized point $x_j$.}
    \ct{Here,} $\varepsilon$ is an additive noise vector with components $\varepsilon_1(x_j)$ and $\varepsilon_2(x_j)$, where $\varepsilon_1(x_j) \sim \mathcal{N}(0, \sigma_1)$ and $\varepsilon_2(x_j) \sim \mathcal{N}(0, \sigma_2)$ are i.i.d. across all $x_j$.
    The standard deviations are defined as $\sigma_1 = \eta \max(|u_{1}^{true}|)$ and $\sigma_2 = \eta \max(|u_{2}^{true}|)$, and 
    the noise level is adjusted by the \ct{user-defined} parameter $\eta$.
    The true solutions $u_1^{true}$ and $u_2^{true}$, shown in \Cref{fig:synthetic_true_solution}, are defined as \ct{normalized versions of {continuous piecewise linear and sine} 
    functions:}
    \begin{equation} \label{eq:synth_true_sol}
    \begin{split}
    	u_1^{true}(x) = \begin{cases}
    		\frac{\sin(2\pi x) - \ct{c_1}}{\Delta_1} & \text{for } x\leq0.5 \\ \frac{-2\pi x + \pi - \ct{c_1}}{\Delta_1} & \text{for } x > 0.5
    	\end{cases}, \\ u^{true}_2(x) = \begin{cases}
    	\frac{-2\pi x + \pi - \ct{c_2}}{\Delta_2} & \text{for } x \leq 0.5 \\ \frac{\sin(2\pi x)- \ct{c_2}}{\Delta_2} & \text{for } x>0.5
        \end{cases},
    \end{split}
    \end{equation}
    {where $\ct{c_1} = -0.536$, $\ct{c_2} = 0.536$ 
    and $\Delta_1 = \Delta_2 = 1.341$ are the \ct{mean and range of the underlying (unnormalized) continuous functions}. 
    Note that \ct{the standard deviations are the same, i.e.,} $\sigma_1 = \sigma_2$, since $\max(|u_{1}^{true}|) = \max(|u_{2}^{true}|)\approx 0.64$.}
    
    {Both model problems share the same} bimodal function for the true diffusion parameter $m_{true}(x)$, {defined as}:
    \begin{align}
        m_{true}(x) = e^{-50(x-0.2)^2} + e^{-50(x-0.8)^2}.
    \end{align}
    This bimodal structure comprises two regions with distinct mixing characteristics, simulating the rapid changes in the diffusion coefficient across different locations within the domain. 
    The exchange term is defined by setting $\omega = 1$ and $d(x) = e^{-10(x-0.3)^2}$. 
    The manufactured solutions $u_1^{true}$ and $u_2^{true}$ are included to compute source terms $f_1(x)$ and $f_2(x)$ to ensure consistency with \Cref{eq:reminder_us_gov_eq}.
    The associated boundary conditions are defined as $u_{1}(0) =0.371$ and $u_{1}(1) = -2.033$ for Dirichlet boundary conditions and $e^{m(x)} \frac{du_2}{dx}(x)\big|_{x=0} = -2.761 $ and $e^{m(x)} \frac{du_2}{dx}(x)\big|_{x=1} = 2.895$ for Neumann boundary conditions.
    
    
    \textbf{Inverse problem approaches.} The solutions $u^{true}_1$ and $u^{true}_{2}$ exhibit critical points, \ct{where $du/dx = 0$,} at $x=0.25$ 
    and $x = 0.75$, \ct{respectively.}
    {This \ct{design choice} significantly impacts the sensitivity of the inverse problem, as discussed in \Cref{sec:sens_for_single_n_mult}.}
    \ct{The data sets $u_{d1}$ and $u_{d2}$ used in the inversion are generated from these true solutions with added noise, as shown in \Cref{eq:synth_data}.}
    \ct{For convenience, we refer to the corresponding inversion experiments as left critical point (LCP) and right critical point (RCP):}
    \begin{enumerate}
        \item Single-Data LCP -- Optimization is done with data $u_{d1}$ only. The forward solution will satisfy \cref{eq:reminder_us_gov_eq} for $i=1$ and the Dirichlet boundary conditions.
        \item Single-Data RCP -- Optimization is done with data set $u_{d2}$ only. The forward solution will satisfy \cref{eq:reminder_us_gov_eq} for $i=2$  and the Neumann boundary conditions.
        \item Double-Data -- Optimization with both $u_{d1}$ and $u_{d2}$ with the format shown in \Cref{eq:cost_functional} where $N=2$. Each forward solution will satisfy its own corresponding forward problem and boundary condition.
    \end{enumerate}
    The numerical results presented in this paper were obtained using hIPPYlib~\citep{VillaPetraGhattas21}, a Python library for inverse problems that implements state-of-the-art scalable adjoint-based algorithms for deterministic and Bayesian inverse problems. 
    It builds on FEniCS~\citep{DupontHoffmanJohnsonEtAl03, LoggMardalWells12} for the discretization of the DEs and on
    PETSc~\citep{BalayBuschelmanGroppEtAl01a,BalayBuschelmanGroppEtAl09} for scalable and efficient linear algebra operations and solvers needed for the solution of 
    DEs.
    The domain is discretized \ct{using $N_x+1$ nodes, yielding $N_x$ uniformly spaced elements.} 
    For the forward and adjoint problems as well as their incremental counterparts, a piecewise-linear finite element discretization is used. 
    The unknown inversion field $m$ uses the same linear finite elements.
    The underlying optimization problem is solved using an inexact Newton-CG method, as detailed in \Cref{alg:method_alg}.
    The code for this problem can be found in the following~\href{https://github.com/aho38/Diffusion-Coefficient-Estimation}{GitHub repository}.

    \subsection{Convergence analysis and robustness testing} \label{sec:robust_test}
    \begin{figure*}
        \centering
        \begin{overpic}[width=0.25\linewidth]{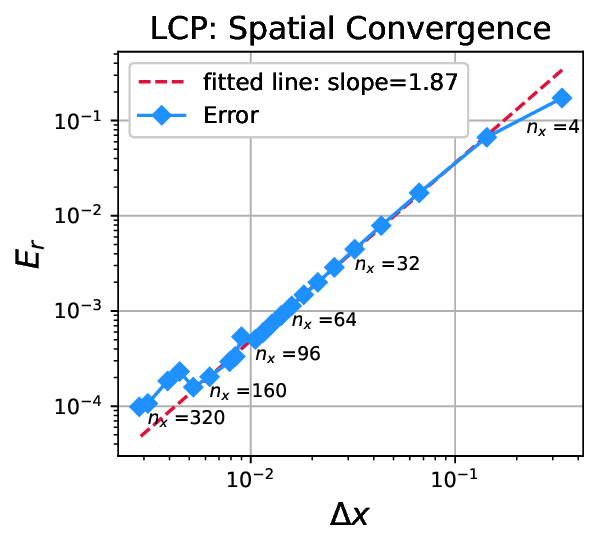}
            \put(15,84.){\footnotesize A.}
        \end{overpic}\hspace{-0.3cm}
        \begin{overpic}[width=0.25\linewidth]{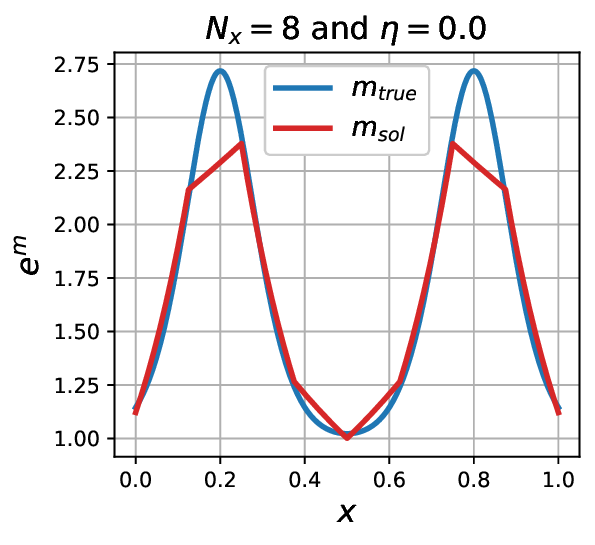}
            \put(25,84.){\footnotesize B.}
        \end{overpic}\hspace{-0.3cm}
        \begin{overpic}[width=0.25\linewidth]{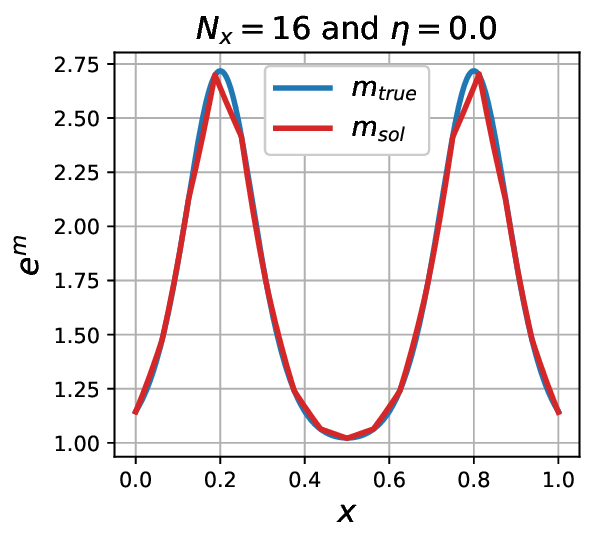}
            \put(24,84.){\footnotesize C.}
        \end{overpic}\hspace{-0.3cm}
        \begin{overpic}[width=0.25\linewidth]{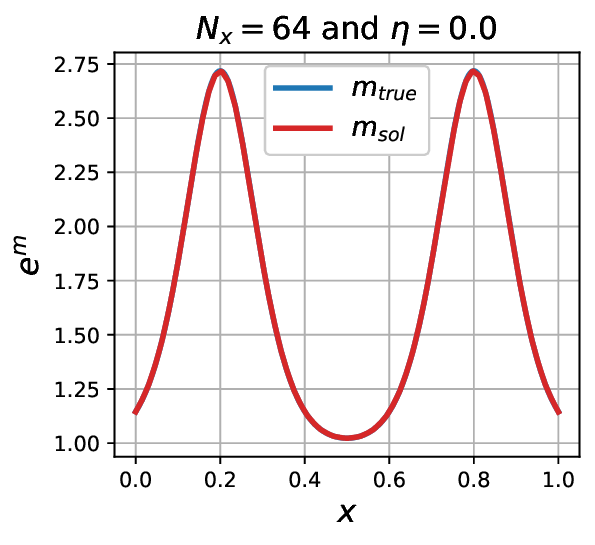}
            \put(24,84.){\footnotesize D.}
        \end{overpic}
        \caption{Panel A. The spatial convergence of the \ct{relative $L_2$ error}, $E_r$ (\cref{eq:error_matric}), for the inferred diffusion coefficient in the absence of noise. As the grid is refined, i.e., $\Delta x \to 0$, second-order convergence is obtained as expected, since a first-order Lagrange polynomial is used as basis functions. 
        Panels B, C, and D show examples of diffusion coefficients inferred with grid resolution,  $N_x = 8, 16, 64$, respectively. }
        \label{fig:spatial_convergence_studies}
    \end{figure*}
    In this section, the spatial convergence and robustness of the inverse problem are evaluated as functions of the grid resolution $N_x$ and the noise level $\eta$, respectively.
    \ct{The spatial convergence study is conducted in a noise-free setting to assess discretization behavior. 
    The remaining synthetic problems are conducted with noise added to the data.}
    The accuracy of the inferred diffusion coefficient, $D(x) = e^{m(x)}$, is assessed with the Single-Data LCP approach using \ct{the relative $L_2$} error:
    \begin{align}\label{eq:error_matric}
        E_r = \ct{\left(\frac{\int_\Omega \left(e^{m_{sol}(x)} - e^{m_{true}(x)}\right)^2\, dx}{\int_\Omega \left(e^{m_{true}(x)}\right)^2\, dx}\right)^{1/2}},
    \end{align}
    \ct{where the domain is $[0,1]$.
    The integral is evaluated using the finite-element discretization and numerical quadrature.
    } \ct{Note that here we impose the correct values of $m(x)$ on the boundaries, so that we only aim to recover $m(x)$ in the interior of the domain.}
    \ct{This ``end-to-end'' error measure reflects the cumulative effect of all stages of the inverse procedure, including the forward and adjoint solves, the optimization procedure, and the finite-element discretization. }

    \textbf{Spatial Convergence.} For the spatial convergence study, 
    noise-free synthetic datasets, i.e., $\eta=0$, and $u_{d1} = u_1^{true}$ were used. The regularization parameter is set to $10^{-12}$, rendering it negligible. 
    The number of elements is varied from $N_x=4$ to $N_x = 352$, and the dependence of error on grid resolution is analyzed.
    
    In Panel A of \Cref{fig:spatial_convergence_studies}, it can be seen that a second-order spatial convergence was achieved, which is consistent with our expectation for using first-order Lagrange basis functions. 
    This suggests that the dominant error source arises from the discretization of the forward problem. 
    Three examples of the inferred diffusion coefficient are shown in Panels B, C, and D for $N_x = 8, 16$, and $64$, respectively, confirming convergence to the ``truth'' in the zero noise regime.
    
\begin{figure*}
        \centering
        \begin{overpic}[width=0.26\linewidth]{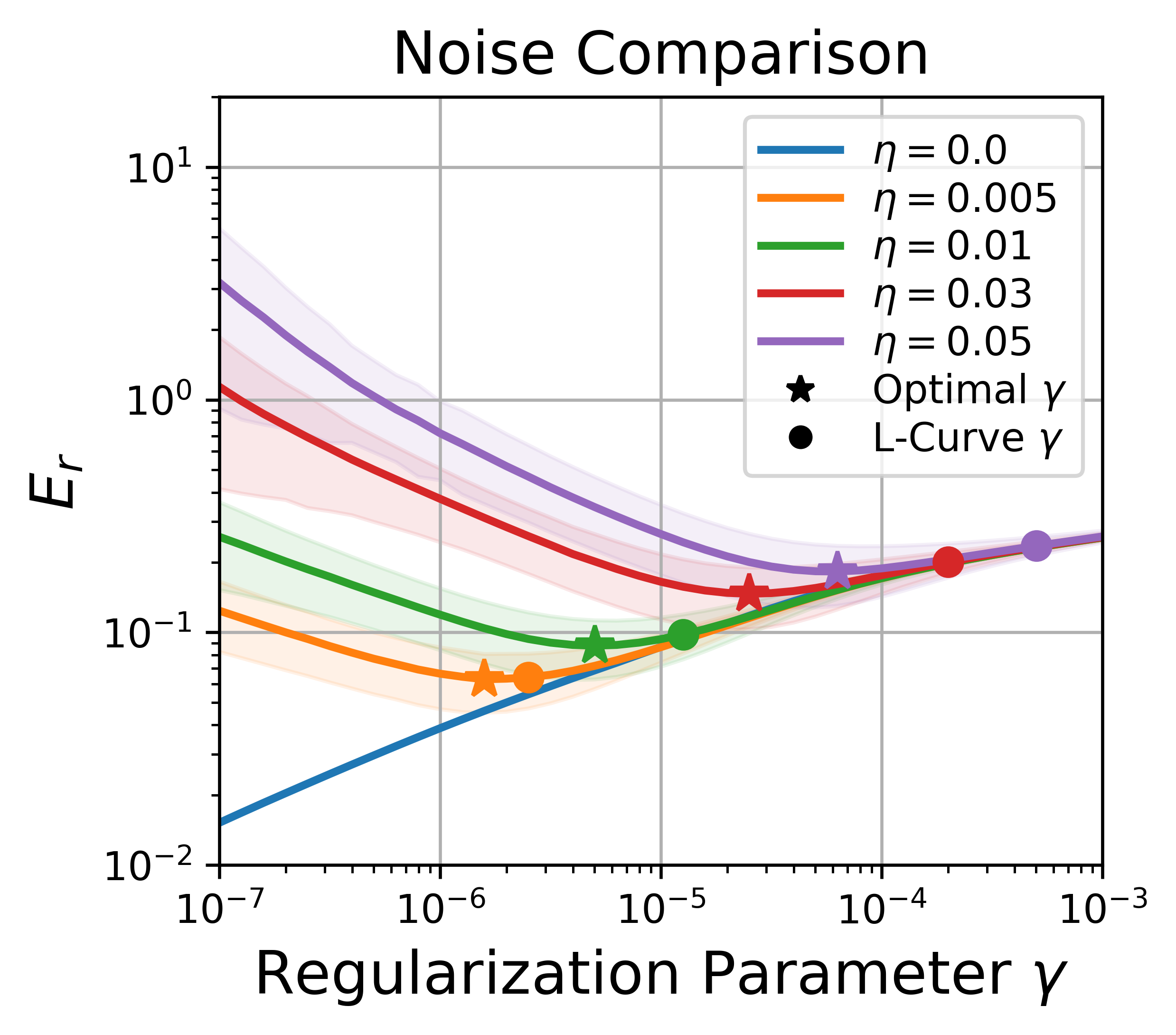}
            \put(23,81.5){\footnotesize A.}
        \end{overpic}\hspace{-0.35cm}
        \begin{overpic}[width=0.248\linewidth]{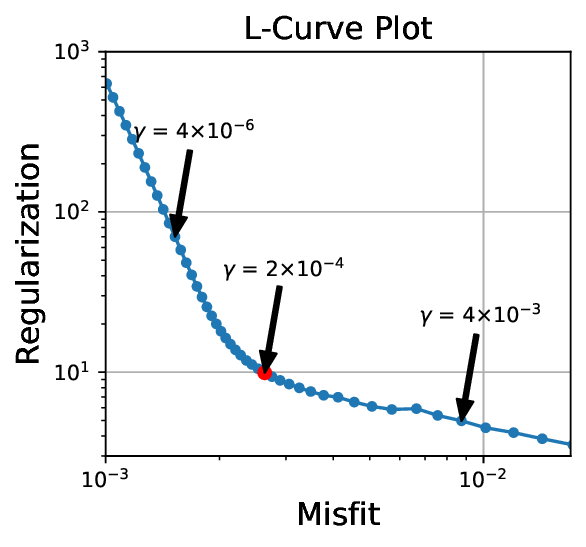}
            \put(30,86.5){\footnotesize B.}
        \end{overpic}\hspace{-0.3cm}
        \begin{overpic}[width=0.248\linewidth]{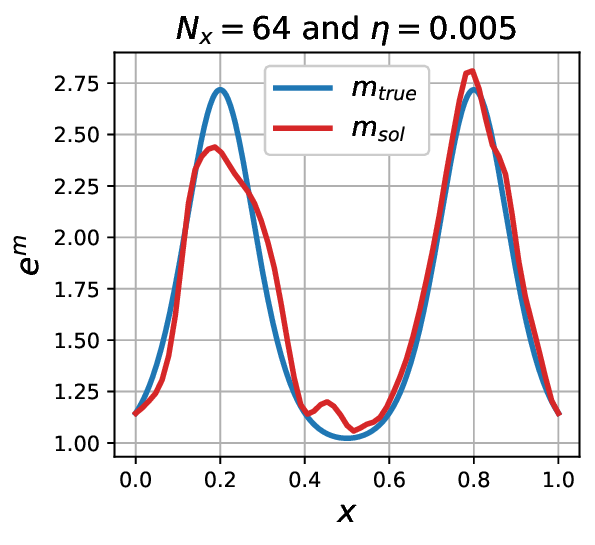}
            \put(20,85.){\footnotesize C.}
        \end{overpic}\hspace{-0.3cm}
        \begin{overpic}[width=0.248\linewidth]{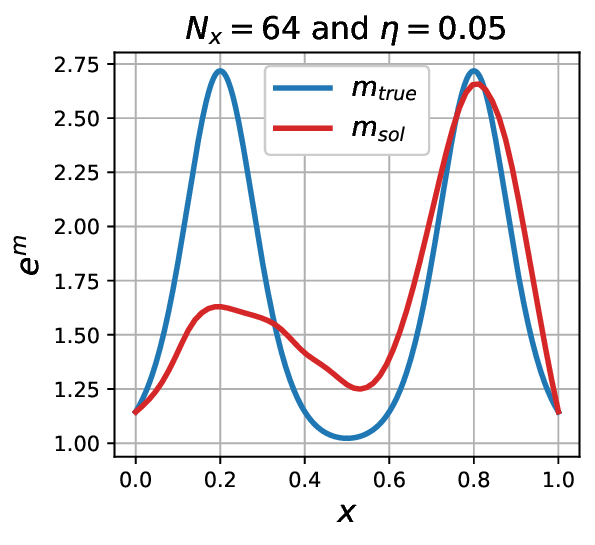}
            \put(20,85.){\footnotesize D.}
        \end{overpic}
        \caption{
        Panel A: The \ct{average of the relative} $L_2$ error, $E_r$ (\cref{eq:error_matric}), 
        versus the regularization parameter for various noise levels. 
        The shading regions indicate one standard deviation over 100 noise realizations.
        Panel B: L-curve-based regularization parameter selection for $\eta=0.03$. 
        Each blue point on the curve represents 
        a different choice of the regularization parameter $\gamma$. 
        The L-curve criterion selects the value at the point of maximum curvature as the ``optimum'' regularization parameter; in this case, $\gamma \approx 2\times10^{-4}$ (shown in red). 
        Panels C and D: \ct{A representative realization of the} true \ct{and} reconstructed diffusion coefficient for two noise levels using $\gamma \approx 2\times 10^{-6}$ for $\eta = 0.005$ in C and $\gamma \approx 8 \times 10^{-5}$ for $\eta = 0.05$ in D. 
        }
        \label{fig:noise_convergence_studies}
    \end{figure*}

    \textbf{Robustness to noise.} To study the robustness of the inverse solver, the accuracy of the inverted diffusion coefficient is analyzed by fixing the grid resolution to $N_x = 64$ and varying the noise level $\eta$ and the regularization parameter $\gamma$.
    The process is repeated 100 times with different noise seeds, and the results \ct{are} shown in \Cref{fig:noise_convergence_studies}. 
    
    Panel A presents the \ct{average of the relative} error of the inverted solution as a function of the regularization parameter $\gamma$. As expected, higher noise levels degrade reconstruction accuracy. However, selecting an appropriate regularization parameter mitigates this effect. 
    The true optimal $\gamma$ that minimizes the error is marked with a star on each curve.
    To approximate the optimal $\gamma$ without prior knowledge, the L-curve method \cite{hansen2001lcurve} is used, giving values shown as dots on each curve.
    
    Panel B shows an example of an L-curve corresponding to $\eta = 0.03$. 
    The trade-off between the magnitude of the regularization and that of the misfit exhibits the characteristic ``L''-shape.
    The chosen $\gamma$, determined by maximizing the curvature of the L-curve, is marked by a red circle and labeled $\gamma = 2 \times 10^{-4}$.
    The selection closely approximates the true optimal $\gamma$, but its effectiveness diminishes as noise increases.
    Nevertheless, the relative error is generally insensitive to small variations in $\gamma$, suggesting a robustness in the choice of the regularization parameter.
    
    Lastly, qualitative assessments are provided in Panels C and D with inverted diffusion coefficients at noise levels $\eta = 0.005$ and $\eta = 0.05$, respectively.
    These results illustrate that increasing noise leads to larger deviations from the true diffusion coefficient, especially in regions \ct{of low sensitivity, as discussed in the next section.} 
    Note that in the absence of noise,  the parameter is reconstructed exactly (\Cref{fig:spatial_convergence_studies}, Panel D). Applying L-curve-selected regularization helps stabilize the reconstruction, even when the noise level is high.
    
    \subsection{Sensitivity analysis for single and multi-data inversion} \label{sec:sens_for_single_n_mult}

        In this section, a qualitative analysis of a single inferred diffusion coefficient is provided. 
    Specifically, the results from the Single-Data LCP approach, with all parameters as described in \Cref{sec:num_setup}, are evaluated.
    The noise level was set to $\eta = 0.03$, the grid resolution to $N_x = 64$, and the regularization parameter to $\gamma=1.994 \times 10^{-4}$, chosen by the $L$-curve method. 
    
\begin{figure*}
        \centering
        \includegraphics[width=\linewidth]{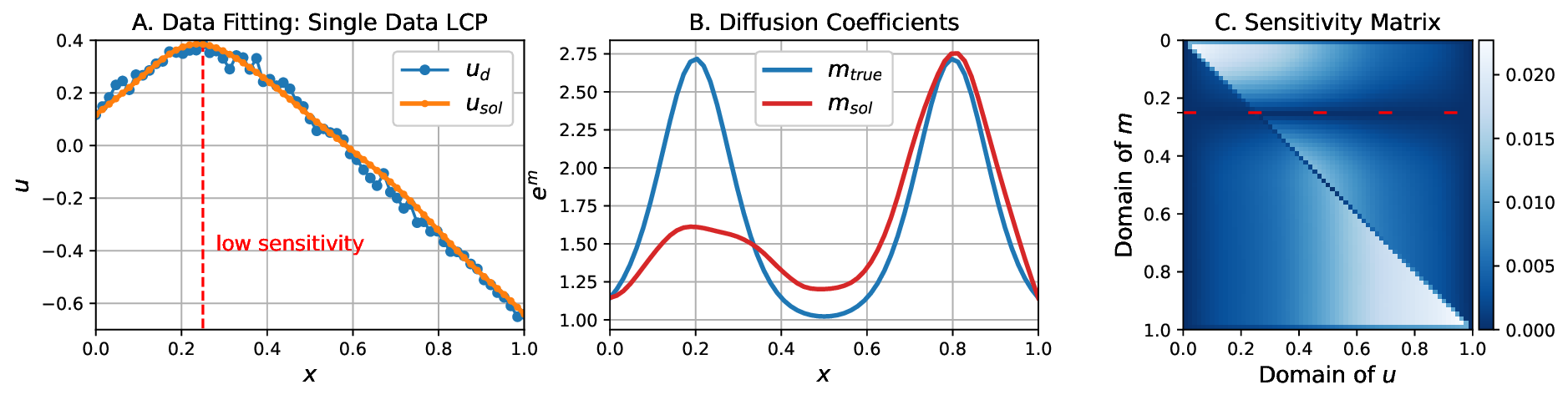}
        \caption{Panel A. 
        Data, $u_d$, in blue, and recovered solution, $u_{sol}$, in orange, from problem 1 in \cref {eq:synth_true_sol}.
        Panel B. The exact (blue) and reconstructed (red) effective diffusion coefficient.
        Panel C. Absolute value of the sensitivity matrix, $\left|\tfrac{\partial u(x_j)}{\partial m(x_i)}\right|$, for $i,j = 1,\dots, \ct{N_x+1}$. The regularization parameter is selected using the L-curve 
        ($\gamma = 1.994 \times 10^{-4}$) and grid resolution is $N_x=64$. 
        The red dashed line in panels A and C highlights the point of lowest sensitivity. 
        }
        \label{fig:flat_u_effect_plot}
    \end{figure*}

    Panel A of \Cref{fig:flat_u_effect_plot} compares the recovered $u$ against data $u_{d1}$ and shows that the region of low sensitivity, indicated with a dashed line, is around $x=0.25$. 
    The solution $u$ exhibits a good fit to the noisy data without overfitting, demonstrating the effectiveness of the regularization.
    Panel B compares the inferred diffusion coefficient, $m_{sol}$, with the true diffusion coefficient, $m_{true}$. 
    While the algorithm effectively reconstructed the second mode, it struggles with the first mode.
    This uneven reconstruction is explored through a sensitivity analysis.
    
    A key challenge in the inverse problem arises from regions where the norm of the derivative of the solution $u$ is small, i.e. $\lVert \nabla u \rVert \approx 0$. 
    In such areas, variations in the diffusion coefficient have a negligible effect as they multiply a small quantity.
    This results in a lower sensitivity $\frac{\partial \textbf{u}}{\partial \textbf{m}}$, and thus a greater susceptibility to error in the inverse solution and potential instabilities in the recovered solution.
    Investigating the right-hand-side of the sensitivity term shown in \Cref{eq:sen_eq_cont}, the following integral is obtained (note that $\mu = m$ as defined above \cref{eq:def_r}):
    \begin{align}\label{eq:sen_eq_rhs}
        \frac{\partial r_k}{\partial \mu_j} = \into e^{m(x)}{\psi_j(x)}\nabla \phi_k(x) \cdot \nabla u(x) dx  = \hat{\textbf{C}}_{j,k}(m,u)
    \end{align}
    where $\hat{\textbf{C}}_{j,k}$ is the $k^{th}$ entry of row $\hat{\textbf{C}}_j = \frac{\partial \textbf{r}}{\partial \mu_j}$ as defined in \Cref{sec:sens_analysis} and \cite{petra2011}. 
    Note that this vector with entries indexed by $k$ is only non-zero when $k$ is near $j$ because finite element basis functions, $\phi_k(x)$ and $\psi_j(x)$, are compactly supported.
    \ct{This property results in a small number of nonzero entries and leads to the banded structure of $\hat{C}$.} 
    These values, however, are negligible when $\nabla u \approx 0$, resulting in the sensitivity obtained from \Cref{eq:sen_eq_cont} to also be negligible.
    As a result, any perturbation on parameter $m$ has little effect on the residual and, therefore, on the solution $u$. 
    More details on the sensitivity and how \Cref{eq:sen_eq_rhs} is derived are discussed in \cite{ho2024inv}.
    
    

    
    Panel C of \Cref{fig:flat_u_effect_plot} shows the absolute value of the sensitivity matrix, $\left|\tfrac{\partial \textbf{u}}{\partial \textbf{m}}\right|$, with the horizontal axis showing the domain of $u$ and the vertical axis the domain of $m$. A low-sensitivity row is observed in the domain of $m$ near $x = 0.25$, indicating a lack of information about the diffusion coefficient at that location.   Correspondingly, the reconstructed diffusion coefficient is poor in this region.
    Note that the near-zero diagonal of Panel C is due to the choice of symmetric basis functions in the discretization process. 
    On the diagonal, the product of $\psi_j(x)$, an even function about $x_j$, and $\nabla\phi_j(x)$, an odd function about $x_j$, yields an odd function.
    When integrated over the domain, this product vanishes in the limit of large $N_x$, leading to zero diagonal entries. 
    However, since $m(x)$ and $\nabla u (x)$ are also present in the integral shown in \Cref{eq:sen_eq_rhs} and are not necessarily constants within each finite element, the integral for the diagonal entries is only approximately zero. 
    This property results in the near-zero diagonal structure observed in Panel C of \Cref{fig:flat_u_effect_plot}.

    \begin{figure*}[b!]
        \centering
        \begin{overpic}[width=0.47\linewidth]{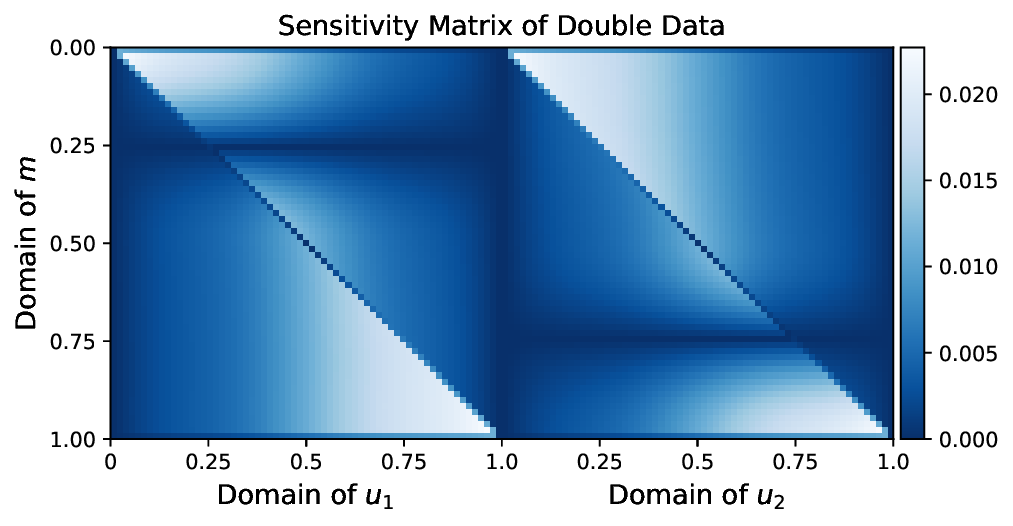}
            \put(21,48.2){\footnotesize A.}
        \end{overpic}
        \begin{overpic}[width=0.515\linewidth]{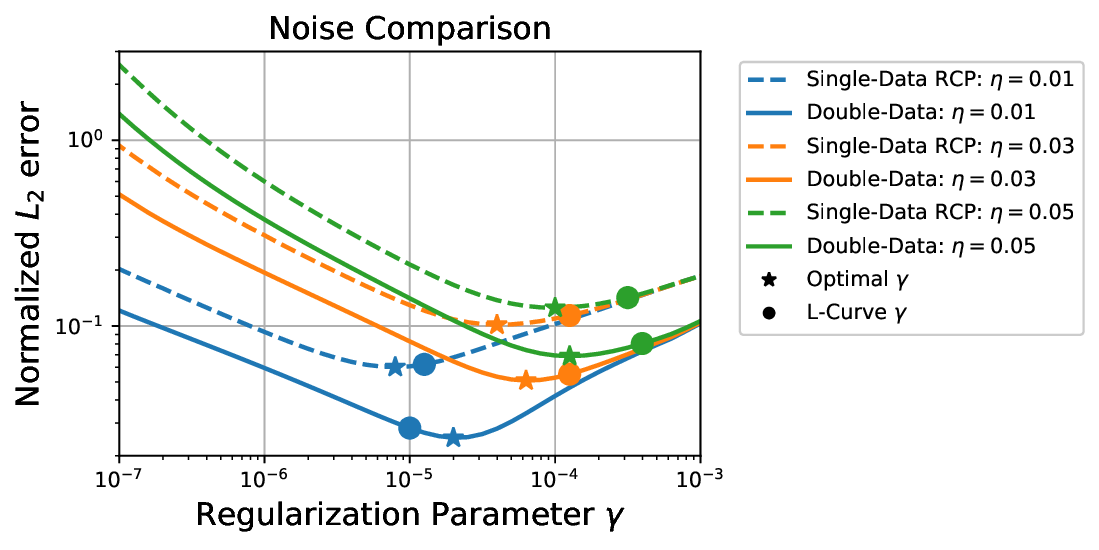}
            \put(19,46.5){\footnotesize B.}
        \end{overpic}
        \caption{Panel A. Sensitivity of $u_1$ and $u_2$ with respect to the shared parameter $m$. Panel B. Relative error of inverted diffusion coefficient with varying regularization parameters at different noise levels, $\eta$. Each noise level is run with 100 different seeds. Double-Data results are compared with Single-Data RCP results, as Single-Data RCP outperforms Single-Data LCP. The location at which $\gamma$ minimizes the relative error of inverted parameter is marked with $\bigstar$ while $\bullet$ is used to indicate the optimal $\gamma$ selected by the $L$-curve.} 
        \label{fig:dual_data_nx_convergence}
    \end{figure*}
    
    \subsection{Comparison of inversion results with single and double data}
    Consider now the Double-Data setup where $N=2$ in \cref{eq:cost_functional}. 
    Both data are utilized during optimization, and other parameters are as described in \Cref{sec:num_setup} with the addition of  $\beta_1 = \beta_2 = 0.5$.
    For spatial convergence, the results are similar to those shown in \Cref{fig:spatial_convergence_studies}, with a second-order convergence.
    For noise convergence, the errors of the inferred diffusion coefficient between Single-Data RCP and Double-Data setups are compared. 
    Note that Single-Data RCP was selected for comparison because it produces a better diffusion coefficient reconstruction overall than Single-Data LCP. 
    Here again, noise is generated for the data using 100 different random seeds, and the results presented are the average relative error of the inferred diffusion coefficient $e^{m(x)}$. 
    
\begin{figure*}
        \centering    
    
        \begin{overpic}[width=0.33\linewidth]{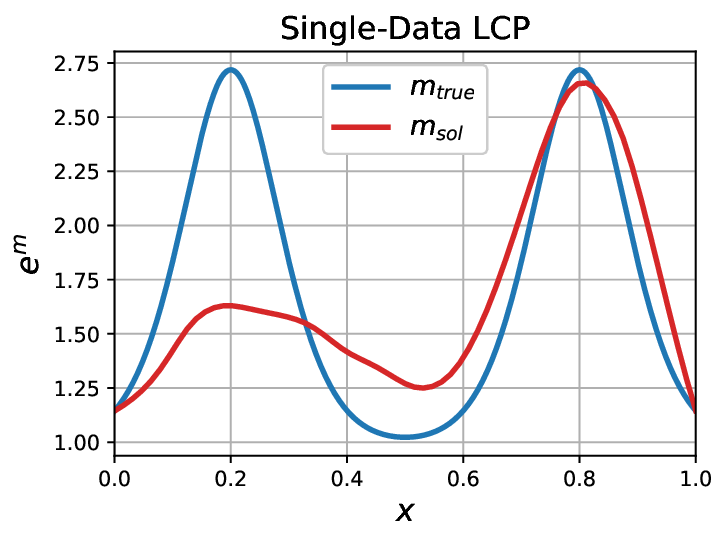}
            \put(31,69.){\footnotesize A.}
        \end{overpic}\hspace{-0.2cm}
        \begin{overpic}[width=0.33\linewidth]{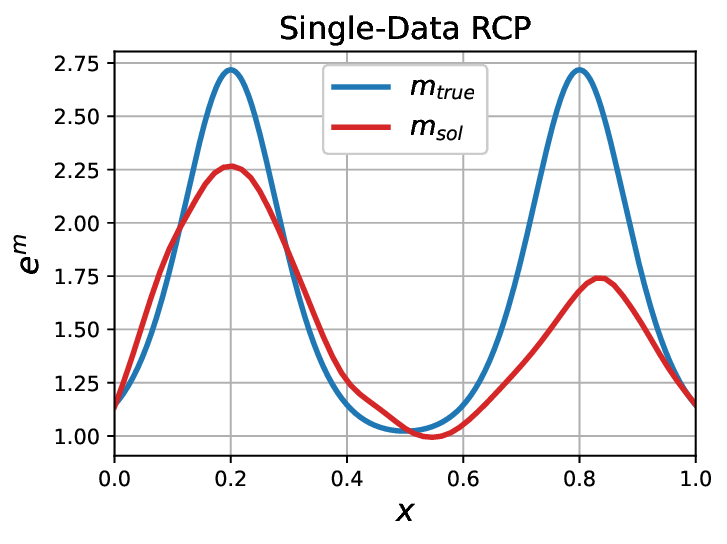}
            \put(31,69.){\footnotesize B.}
        \end{overpic}\hspace{-0.2cm}
        \begin{overpic}[width=0.33\linewidth]{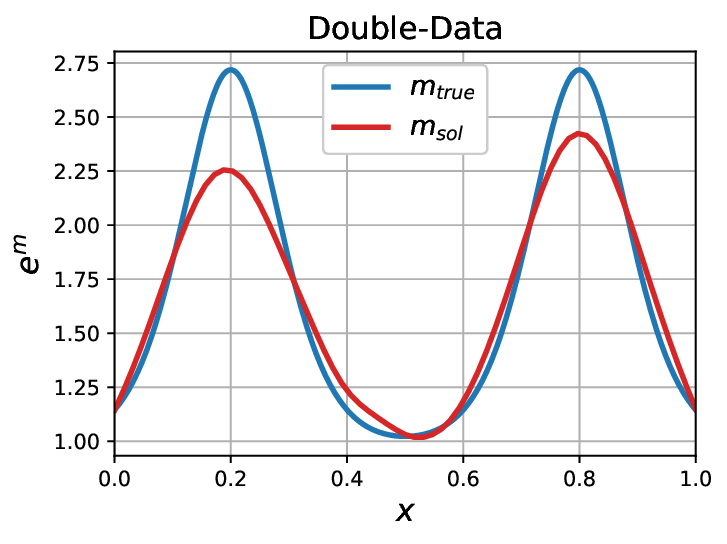}
            \put(34,69.){\footnotesize C.}
        \end{overpic}
        \caption{
        Inversion results with Single-Data (Panel A for LCP and Panel B for RCP) and Double-Data (Panel C). 
        These results were obtained with a grid resolution of $N_x = 64$, noise level of $\eta = 0.05$, and  
        regularization parameter $\gamma$ chosen using the $L$-curve criterion. 
        }
        \label{fig:diff-recovery-single-double}
    \end{figure*}

    The comparison of Single-Data to Double-Data is shown in \Cref{fig:dual_data_nx_convergence}. 
    Panel A shows the stacked sensitivity matrix for $u_1$ and $u_2$ with respect to $m$.
    The matrix shows that the sensitivity in the domain of $m$ is higher for at least part of $u_1$ or $u_2$, yielding additional information when optimizing with both data concurrently.
    Panel B shows the effect of noise on the error of the inferred diffusion coefficient with various regularization parameters. 
    The results show that, across all regularization parameters, the error for Double-Data is lower than that for Single-Data RCP.
    The overall trend of larger regularization parameters for larger noise levels remains consistent for both Single-Data RCP and Double-Data. 
    For demonstration purposes, the inferred diffusion coefficient for Single-Data LCP, Single-Data RCP, and Double-Data are compared in \Cref{fig:diff-recovery-single-double}, in Panels A, B, and C, respectively.
    Note that the qualitative difference in the inversion of the two Single-Data cases can be attributed to the exchange term characterized by $d(x)$.
    As defined in \Cref{sec:num_setup}, $d(x)$ is a Gaussian function with a larger value near the first mode and a smaller value near the second mode, resulting in a more dominant exchange term in the first half of the domain.
    The stronger exchange term reduces the relative influence of diffusion in that region, leading to poorer inversion or reconstruction of the diffusion coefficient. 
    However, the Double-Data reconstruction is more accurate than even the Single-Data RCP.
    
    \subsection{Hessian spectrum and computation cost analysis} \label{sec:comp_analysis}

    \begin{figure}
        \centering
        \includegraphics[width=1.\linewidth]{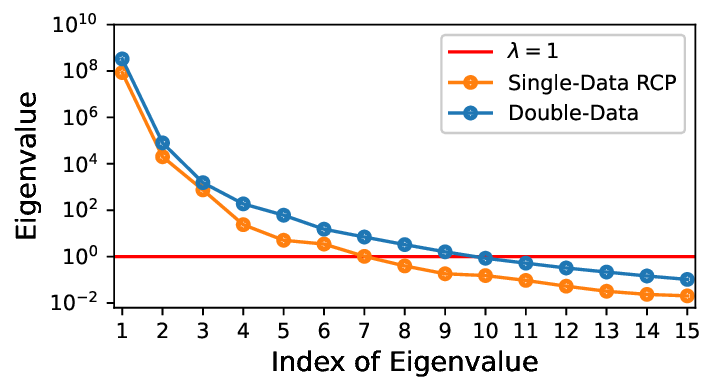}
        \caption{
         Log-linear plot of the spectrum of the prior-preconditioned data misfit Hessian for the inverse problem using double-data. 
         The plot shows the 15 largest eigenvalues, with the Double-Data truncated at $0.1$.
         }
        \label{fig:spectrum}
    \end{figure}
    \begin{table}[width=\linewidth,cols=4,pos=h]
        \caption{
        Average number of Newton and CG iterations for the inexact Newton method for the inverse problem with Single-Data and Double-Data. We obtained these results from 100 runs, each initialized with a different random seed.
        }
        \begin{tabular*}{\tblwidth}{@{} L|L|L|L@{} }
         & \makecell{Avg. Newton \\ Iteration}  & \makecell{Avg. CG \\ (Fixed tol)} & \makecell{Avg. CG \\ (Adapt. tol)} \\ \hline 
        Single-Data LCP & $3.00$ & $9.52$ & $5.96$ \\ \hline
        Single-Data RCP & $4.02$ & $9.00$ & $5.19$ \\ \hline
        Double-Data & $4.82$ & $14.21$ & $7.17$ \\ 
        \end{tabular*}
        \label{tb:iter_num}
    \end{table}

    This section further investigates the influence of diverse data used in the inverse problem by looking at the spectrum of the preconditioned reduced-Hessian misfit, defined as $\textbf{H}_p$ in \Cref{eq:precond_hessian}.
    \Cref{fig:spectrum} shows the dominant eigenvalues of $\textbf{H}_p$ for the Single-Data RCP and Double-data setups.
    In \Cref{fig:spectrum}, a value of $\lambda = 1$ is marked, for which the eigenvalues below that have significantly less impact due to the error bound when approximating the preconditioned Hessian misfit using the Sherman-Morrison-Woodbury formula \cite{Ghattas_Willcox_2021,bui2013computational}. 
    These results reveal that the eigenvalues of the double-data Hessian misfit decay at a slightly slower rate than those of the single-data case, confirming that the double-data is more informative and leads to better parameter inversion.
    Note that, for both Single-Data RCP and Double-Data setups, the rapid decay of the eigenvalues of  $\textbf{H}_p$ is due to the smoothing nature of the DE operator, which essentially reflects the ill-posedness of the inverse problems. 
    From an optimization point of view, this means that the optimization landscape exhibits a degree of "flatness" along directions associated with small eigenvalues~\cite{bui2012analysis}. 
    \ct{This can cause a stopping criterion based on gradient norm, step size, or objective decrease to be met prematurely, resulting in a poorer parameter recovery.} 
    Here, the spectral properties of the Hessian misfit term were used only to better assess the quality of the inversion results. 
    In general, one can also use them to construct low-rank approximations and accelerate the optimization procedure \cite{petra2012, Ghattas_Willcox_2021, bui2013computational, VillaPetraGhattas16}. 
    
    When the decay of eigenvalues is slower, the conjugate gradient (CG) takes longer to converge when solving the Newton system, \Cref{eq:newt_linear_system}, when the termination tolerance is fixed. 
     \citet[Chapter 5, Section 2]{NoceWrig06} gives an upper bound of the error for the $k^{th}$ CG iteration as a function of the $(\ct{N_x+1}-k)^{th}$ largest eigenvalue of the matrix of the linear system (of size $\ct{N_x+1 \times N_x+1}$). 
    If \textbf{H} has eigenvalues of $\lambda_1 \leq \lambda_2 \leq \hdots \leq \lambda_{\ct{N_x+1}}$, giving:
    \begin{align}\label{eq:eig_cg_bound}
        \lVert \hat{\textbf{m}}_{k+1} - \hat{\textbf{m}}^*\rVert_H^2 \leq \left(\frac{\lambda_{\ct{N_x+1}-k} -\lambda_1}{\lambda_{\ct{N_x+1}-k} +\lambda_1}\right)^2 \lVert\hat{\textbf{m}}_0 - \hat{\textbf{m}}^*\rVert_H^2,
    \end{align}
    where the norm is $\lVert\hat{\textbf{m}}\rVert_H = \hat{\textbf{m}}^\intercal \textbf{H}\hat{\textbf{m}}$.
    By computing the eigenvalue coefficient in the upper bound of \Cref{eq:eig_cg_bound}, the expected number of CG iterations needed for the norm of the residual to drop below tolerance can be approximated. 
    The estimated number of CG iterations is $15$ for Double-Data and $10$ for Single-Data (for both LCP and RCP).
    The average number of experimental CG iterations is shown in \Cref{tb:iter_num}, where the results are consistent with theoretical expectations. 
    Note that this is an approximation without accounting for the norm of $\lVert \hat{m}_0 - \hat{m}^*\rVert^2_H$ in \Cref{eq:eig_cg_bound}. 
    In practice, adaptive tolerance improves the computational efficiency,  {\Cref{eq:tolcg}} \cite{Ghattas_Willcox_2021,dembo1982inexact,eisenstat1996choosing}.
    The adaptive tolerance allows faster CG termination in the first few iterations of Newton's method, resulting in a higher overall computational efficiency, as shown in the third column of \Cref{tb:iter_num}.
    Since Newton's system is solved iteratively using CG with an adaptive tolerance, the resulting \ct{Newton search} direction $\hat{\textbf{m}}$ is only an approximation.
    The inexactness gives the name \emph{inexact Newton-CG}.
    The stopping criteria for Newton's method are a tolerance on the norm of the gradient or on the decrease of the cost functional shown in  \cref{eq:cost_functional}.
    
    To quantify computational costs, the number of differential equation solves is used as a metric.
    Note that this metric quantifies the algorithm's computational burden and is independent of hardware.
    As shown in \Cref{eq:state_eq_general,eq:adj_eq_general}, there are two DEs solved at each Newton iteration, the state equation and the adjoint equation. 
    The incremental state and the incremental adjoint equation shown in \Cref{eq:inc_state_eq_general,eq:inc_adj_eq_general}, respectively, are solved at each CG iteration. 
    Therefore, for $N$ data sets (see \Cref{eq:cost_functional}), the total number of DE solves is $2N \times (N_{\mathrm{new}} + N_{\mathrm{CG}}),$ where $N_{\mathrm{new}}$ is the number of Newton iterations and $N_{\mathrm{CG}}$ is the total number of CG iterations accumulated over all Newton iterations.
    The optimizations for Single-Data LCP and RCP require about $44$ and $52$ DE solves, respectively. 
    For Double-Data optimization, the number of DE solves is approximately $163$. 
    This latter number exceeds the sum of the two Single-Data setups because the Double-Data setup contains more information about the unknown parameter. 
    This leads to an increased numerical rank of the Hessian, and therefore to a higher number of CG iterations and, consequently, to more DE solves~\cite{Ghattas_Willcox_2021}. 
    It also yields a more accurate reconstruction, as noted earlier.

    \subsection{PINNs vs. INCG} \label{sec:pinn_comparison}
    \begin{figure*}
        \centering
        \includegraphics[width=\linewidth]{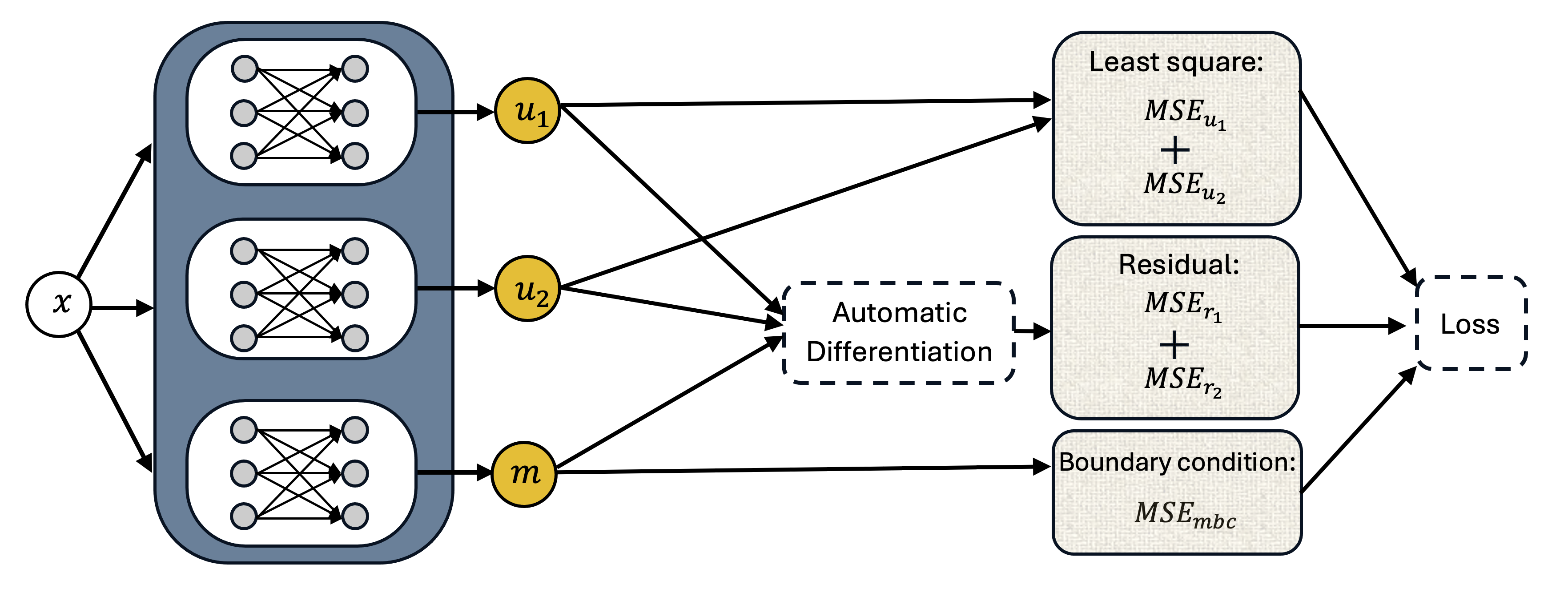}
        \caption{The schematic of the variation of PINNs used for the comparison testing. This setup is an extension of what was presented in \cite{he2021physics, jiang2023practical}. The input spatial feature flows through multi-layer perceptrons, which output the parameters $m$, $u_1$, and $u_2$. Then, the $u_i$ are used to compute a least-squares value against data, as well as the residual after differentiation using auto-differentiation, along with parameter $m$. The boundary condition of $m$ is also assumed to be known, and its loss at the boundary is computed.
        }
        \label{fig:PINN_schematics}
    \end{figure*}
    Given the recent development of physics-informed neural networks (PINNs) for solving and inferring DEs, it is natural to compare their performance with the classical adjoint-based approach. 
    In this section, the present inexact Newton-CG method is abbreviated INCG.
    A recent study by \cite{grossmann2024can} compared PINNs and FEM for data-driven PDE solutions. 
    The present study builds on Grossmann's work. It compares PINNs and INCG, focusing on their performance in inferring a spatially varying diffusion coefficient in a screened Poisson equation from two combined data sets. 
    
    \textbf{Problem formulation.} The PINNs optimization method and architecture presented in \cite{he2021physics,grossmann2024can} are used as benchmarks in this comparison, and the architecture is illustrated in \Cref{fig:PINN_schematics}. 
    Note that this architecture is an adaptation of what is used in \cite{he2021physics, jiang2023practical,rafati2018deep} to accommodate multiple observation data with multiple governing equations.
    The architectures of all sub-networks are identical, but they do not share the same parameters.
    The network takes spatial coordinates as inputs and uses the sub-networks to output solutions $u_1(x)$ and $u_2(x)$ that represent {approximate solutions} to the strong-form of the DEs \cref{eq:reminder_us_gov_eq}. 
    A separate sub-network maps spatial coordinates to the parameter $m(x)$ that is shared by the two governing equations.
    These three components are used to compute the loss, or objective, function 
    \begin{align}
        \text{Loss} = \text{MSE}_{mbc}+ \sum_{i=1}^2\left(\text{MSE}_{u_i} + \text{MSE}_{r_i} \right)
    \end{align}
    where $i$ is the index of the data set used and the equation they satisfy.
    Here, $\text{MSE}_{u_i}$ is the least-square objective term defined as 
    \begin{align} \label{eq:mse_misfit_pinn}
        \text{MSE}_{u_i} = \frac{1}{\ct{N_x+1}}\sum_{j=1}^{\ct{N_x+1}}|u_i(x_j) - u_{di}(x_j)|^2
    \end{align}
    with $u_i(x_j)$ the output of the neural network and $u_{di}(x_j)$ the data points within the domain. 
    The residual loss, $\text{MSE}_{r_i}$, can be expressed as 
    \begin{align}\label{eq:mse_r}
        \text{MSE}_{r_i} = \frac{1}{N_r} \sum_{j=1}^{N_r}\left|r(u_i(x_j), m(x_j))\right|^2
    \end{align}
    where $N_r$ is the number of points sampled using Latin Hypercube Sampling \cite{stein1987large} to evaluate the residuals of \cref{eq:reminder_us_gov_eq}.
    The residual term $r(\cdot)$ is evaluated using auto-differentiation \cite{baydin2018automatic}. Note that in INCG, the residual is zero, up to the numerical accuracy of the forward solver. 
    In contrast, in PINNs, the DEs may have a non-zero residual to improve the fit to the data. 
    Lastly, as was the case in the synthetic problem, the boundary values, $m_k$, of the diffusion coefficient are assumed to be known, and a penalty term is added:
    \begin{align}
        \text{MSE}_{mbc} = \frac{1}{N_m}\sum_{k=1}^{N_m} \left| m(x_k) - m_k \right|\ct{^2}.
    \end{align}
    Since the domain of interest is one-dimensional, namely $\Omega \in [0,1]$, the boundary of the diffusion parameter consists of the endpoints of the domain, i.e., $N_m = 2$. 
    
    \textbf{Numerical experiment setup.} 
    As described in \cref{sec:num_setup}, all observation data $u_{di}$ shown in \Cref{eq:mse_misfit_pinn} are generated from an analytical function, and the noise is generated separately, ensuring that the data for INCG and PINNs are identical. 
    The parameter $N_r$ shown in \Cref{eq:mse_r} is set to $N_r=1000$. 
    Note that this setup provides PINNs with more information than INCG, since the latter evaluates the differential equation only at each data point. 
    The code for all PINNs setups is given in this \href{https://github.com/aho38/Pytorch-PINN-Marine-Lake}{GitHub repository} and is written in PyTorch \cite{paszke2019pytorch}.
    
    \begin{figure*}
        \centering
        \begin{overpic}[width=1.0\linewidth]{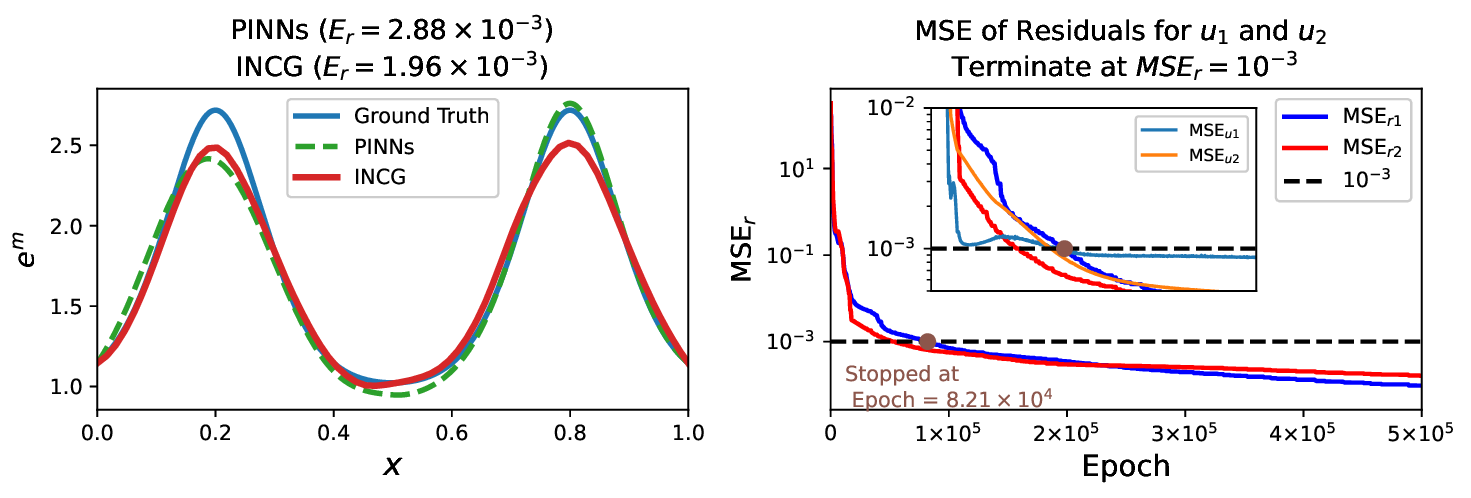}
            \put(8,29.2){\normalsize A.}
            \put(58,29.2){\normalsize B.}
        \end{overpic}
        \caption{Panel A. 
        Parameter inversion comparison between PINNs and INCG. 
        The normalized $L_2$ error, $E_r$ (\cref{eq:error_matric}), of both reconstructions is given.
        Panel B. 
        MSE$_r$ of $u_1(x)$ and $u_2(x)$. The stopping criteria are that both residual losses be less than $10^{-3}$, and the algorithm stops after $1.57 \times 10^{5}$ epochs. In the embedded plot, both misfit losses reached $10^{-3}$ before the residual, indicating that the stopping criterion on the residual of the governing equation also corresponds to a small misfit.
         }
        \label{fig:PINN_result_compare}
    \end{figure*}
    
    \textbf{Discussion.} 
    The inferred diffusion coefficients for both INCG and PINNs are shown in \Cref{fig:PINN_result_compare}.
    Qualitatively, both approaches yield good reconstructions and capture the bimodal shape of the diffusion coefficient. 
    The normalized $L_2$ error was found to be $E_r = 2.88\times 10^{-3}$ for PINNs and $E_r = 1.96\times 10^{-3}$ for INCG, showing that both methods can give a good solution to an inverse problem governed by the screened Poisson equation shown in \Cref{eq:helm_main}.
    
    Physics-informed neural networks 
    offer a straightforward model construction, avoiding the need to formulate a Lagrangian as required by the traditional constraint optimization framework. 
    Auto-differentiation enables the network to learn the mapping from the spatial domain to the observed data, with the residual of the differential equation serving as a regularization term that enforces the underlying physics. 
    However, the black-box nature of neural networks introduces challenges in architecture selection and hyperparameter tuning. 
    Ablation studies indicate that an architecture comprising fully connected layers with 50 neurons and sigmoid activations performs best. While true relative error can be computed in synthetic problems with known solutions to help adjust hyperparameters, only the convergence of the loss function serves as an indicator 
    of successful learning when working with real-world data.
    
    In contrast, INCG employs an optimization framework that ensures the governing equation is satisfied by solving the forward problem using finite element methods. 
    Moreover, INCG leverages the adjoint method to compute gradients efficiently. 
    While this approach requires familiarity with constrained optimization and adjoint methods, it eliminates the need for architecture selection and significantly reduces the complexity of hyperparameter tuning inherent to PINNs. 
    Although formulating the Lagrangian and implementing the adjoint method can introduce complexity in the initial setup, this is mitigated by using a modular version of INCG with hIPPYlib, thereby making the process accessible to users. The trade-off between the two methods lies in the upfront complexity of INCG versus the experimental effort required for architecture selection and hyperparameter tuning in PINNs.
    
    Directly comparing the performance of PINNs and INCG on an inverse problem is challenging due to differences in their implementation and hardware requirements, as noted in \cite{grossmann2024can}.  
    Therefore, this comparison study only provides an approximate evaluation of the runtime for both methods under typical usage.
    For this experiment, the neural network was trained on an Nvidia P100 PCIe 40 GB GPU, and the INCG optimization was performed on the Apple M2 Ultra CPU. 
    The training time was measured until the residual error of \Cref{eq:mse_r} reached $10^{-3}$ as shown in Panel B of \Cref{fig:PINN_result_compare}, which required approximately $2.0\times 10^{3}$  seconds. 
    In contrast, the runtime of INCG with the convergence criteria described in the previous sections was found to be $2.0\times 10^{2}$ seconds, required to optimize $100$ times with different regularization parameters to construct the L-curve. 
    Thus, the PINN requires roughly 10 times as much computational time as INCG. 
    Note that Panel B of \Cref{fig:dual_data_nx_convergence} shows a weak dependence of the error on the regularization parameter, suggesting that a coarser L-curve could suffice to select an appropriate regularization parameter, thus further reducing INCG's runtime.
    
    In summary, both methods provided good solutions to this inverse problem. 
    Similar results were obtained with both methods when fitting a single dataset exhibiting low sensitivity in parts of the domain. 
    While INCG requires familiarity with adjoint methods, it offers advantages in terms of convergence analysis and reduced hyperparameter tuning, particularly with our modular FEniCS implementation. Conversely, PINNs are easier to set up initially, owing to well-documented libraries such as PyTorch and TensorFlow.
    However, PINNs can be computationally expensive to train.
    Note that this comparison is at a relatively small scale, and the differences may become more pronounced in large-scale problems.

\section{Application to a marine lake} \label{sec:marine_application}

    \subsection{Application setup} \label{sec:app_setup}
    \begin{figure}
        \centering
        \begin{overpic}[width=1\linewidth]{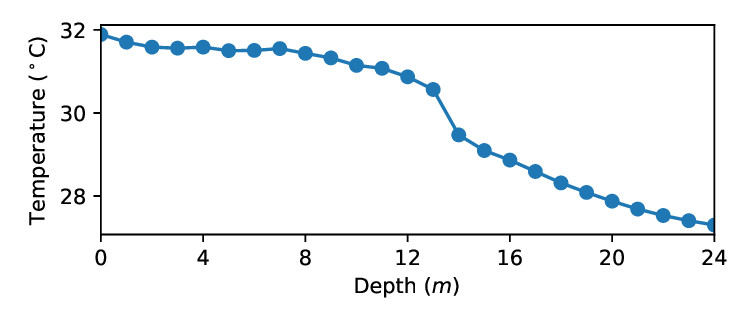}
            \put(16,16){\normalsize A.}
        \end{overpic}\vspace{-0.2cm}
        \begin{overpic}[width=1\linewidth]{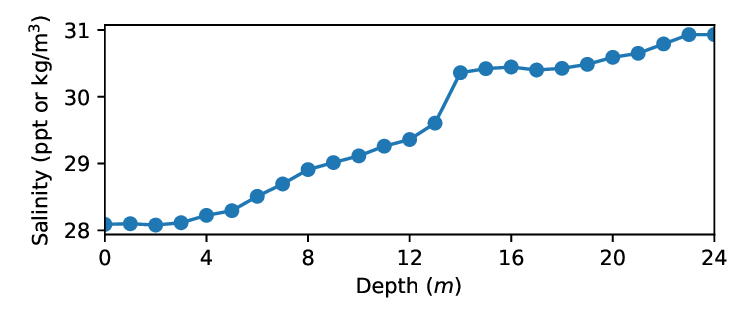}
            \put(16,35){\normalsize B.}
        \end{overpic}\vspace{-0.6cm}
        \caption{The  
        temperature (Panel A) and salinity (Panel B) profiles, averaged over two locations of the lake (east and west sides), on May 13, 2013. 
        }
        \label{fig:real_lake_data}
    \end{figure}
    \begin{table}[width=\linewidth,cols=4,pos=h]
        \begin{tabular}{c c c c}\hline
             Parameter & Unit & Temperature & Salinity \\  \hline
             $\omega$ & 1$/\text{s}$& $5.5\times 10^{-5}$ & $5.5\times 10^{-5}$ \\
             $d(x)$ & - & $\frac{e^{-20(x - 7)^2}}{\into e^{-20(x - 7)^2}\, dx}$ & $\frac{e^{-20(x - 7)^2}}{\into e^{-20(x - 7)^2}\, dx}$ \\
             $I_0$ & $^\circ \text{C}$ & $1.74\times 10^{-6}$ & $0$ \\
             $I(x)$ & - & $\frac{e^{-10x}}{\int_\Omega e^{-10x}\, dx}$ & $0$ \\
             $u_{oce}$ & $^\circ \text{C}$ or $\text{kg}/\text{m}^3$ & $29.0$ & $27.8$ \\
        \end{tabular}
        \caption{
        Parameter values used in the numerical simulations. 
        The rate of inflow ($\omega$) and ocean value ($u_{oce}$) are approximated using the information provided in \citet{blanchette2020marine}. 
        The magnitude of the solar irradiance is denoted with $I_0$ and is obtained from Global Solar Atlas \cite{globalsolaratlas}. 
        The attenuation is captured by $I(x)$.
        }
        \label{tab:real_data_params}
    \end{table}
    
    In this section, the adjoint-based inverse-problem framework is applied to what initially motivated this study: the dynamics of stratified marine lakes, in particular those of Jellyfish Lake (also known as Ongeim'l Tketau), located on Mecherchar Island in Palau \cite{dawson2001jellyfish,n2003geographic}.
    Measurements of temperature and salinity were taken at two locations within the lake on the east and west sides at 1-meter depth intervals down to 24 meters. 
    The dataset used for inversion was obtained by averaging the profiles over the two locations \cite{blanchette2020marine}. 
    Using this dataset, the objective was to characterize mixing within the lake, a dominant physical mechanism that is difficult to measure directly.
    
    
    The 
    dynamics of temperature and salinity profiles are modeled by the screened-Poisson equation shown in \Cref{eq:helm_main} and derived in \cite{blanchette2020marine}. 
    The parameter to infer is the effective diffusion coefficient, $D(x)=D_0 e^{m(x)}$, which characterizes vertical mixing and is assumed to mix temperature and salinity at the same rate, since large-scale turbulent mixing dominates molecular diffusion.
    All parameters of \Cref{eq:helm_main}, except $D(x)$, are either obtained as direct measurements or independently modeled based on existing knowledge of the lake \cite{blanchette2020marine}; they are summarized in \Cref{tab:real_data_params}.
    The exchange rate $\omega$ and distribution function $d(x)$ are common to both temperature and salinity. 
    The distribution function is modeled as a Gaussian centered at $7$ meters below the surface, based on direct observations of tunnels at that depth. 
    Recall that $d(x)$ is averaged over the domain considered.
    The forcing term, $f(x)$, is assumed to be $0$ for salinity as no salt or fresh water is added or subtracted within the domain.
    For temperature, solar forcing is modeled as $f(x) = I_0\, I(x)$, where $I_0$ is the solar irradiance magnitude (value obtained from \cite{globalsolaratlas}) and $I(x)$ describes attenuation with depth. The attenuation function $I(x)$ is modeled as a decaying exponential, with the majority of the light absorbed within the top 10 meters of the water column \cite{jerlov2076chp3}. 
    
    
    As part of the inverse problem setup, boundary conditions are specified for each dataset. The temperature values at the lake surface and bottom are set to match the observed data.
    For salinity, the surface flux includes rainwater and evaporation effects and is used as a Neumann boundary condition at the top boundary.
    Since the rainwater effect dominates evaporation, only the surface freshwater flux from rain is considered, thereby reducing the surface salt concentration through dilution.
    This effect is modeled as $R(u_{\text{surf}} - u_{\text{rain}})$ where $R$ is the rain water volume flux per surface area (m/s), $u_{\text{surf}}$ is salt concentration of the water at the surface, and $u_{\text{rain}}$ is the salt concentration of the rain water, set to $u_{\text{rain}} = 0$. 
    The surface flux is computed as the product of the average annual rate of rainfall and the salt concentration at the surface, yielding $D(0)\nabla u(0)= R \ u_{\text{surf}} = - 10^{-7}$m/s $\times \ 28$kg/m$^3$ \cite{dataset_precipitation2025,dataset_precipitation_2_2025}.
    Note that the magnitude of this flux is very small, resulting in a near-zero derivative condition at the surface.
    The flux at the bottom is not prescribed directly, but instead is inferred using the data compatibility condition described in the next subsection.
    
    Measured temperature and salinity data used in the inversion are shown in \Cref{fig:real_lake_data}. 
    To convert \Cref{eq:helm_main} to the nondimensional form shown in \Cref{eq:helm_nondim}, both temperature and salinity were shifted by subtracting the ocean values, $u_{oce}$, and then scaled by $U=\max(u) - \min(u)$, computed by measured values, so that the nondimensional variables are $(u - u_{oce})/U$ as shown in \Cref{sec:forward_problem_introduction}.
    The temperature and salinity of water influx from tunnels were $29^\circ$C and $27.8$ kg/m$^3$, respectively, based on empirical measurements; their scaling values $U$ were $4.59^\circ$C and $2.85$ kg/m$^3$, respectively.
    These properties of the inflow, $u_{oce}$, differ from typical ocean water due to dilution with groundwater and heat exchange during transit through the subterranean tunnels \cite{blanchette2020marine}.
    
    The adjoint-based inverse algorithm is used to iteratively estimate the effective diffusion coefficient by minimizing the objective functional given in \Cref{eq:cost_functional}, in which real temperature and salinity data are first used separately and then simultaneously. 
    To address the ill-posedness of the inverse problem, the regularization parameter is selected using the L-curve criterion (approximately $\gamma=2.51 \times 10^{-5}$). 
    
    To assess the expected accuracy of the inferred diffusion coefficient, the data noise level is estimated from both instrumental error and measured horizontal variations in temperature and salinity across the lake.
    Data were collected using a Hydrolab Quanta water quality meter, which reports temperature accuracy of $\pm 0.2^\circ$C and salinity accuracy of $\pm 1\%$ of the reading, with a resolution of $\pm 0.01$ on the Practical Salinity Scale (PSS) or kg/m$^3$ in numerical units.
    Here, ``$\pm 1$ count'' corresponds to one increment in the instrument's least significant digit, which for this meter is $0.01$ (i.e., $\pm 1$ PSS or kg/m$^3$).
    To approximate the instrumental noise, we interpret the manufacturer's reported accuracy as corresponding to a $95\%$ confidence interval.
    To quantify variations across the lake, measurements were taken monthly \ct{from 2013 to 2016} at two locations (East and West), and the associated noise was approximated from the observed differences. 
    
    \ct{The results were found to be depth-dependent, with the greatest variations observed at the surface, where the mean temperature difference between east and west measurements was $0.19^\circ$C.
    Variations decreased with depth, dropping to $0.10^\circ$C at 11 meters below the surface and $0.002^\circ$C at 24 meters below the surface. Averaged over the depth of the lake, the mean temperature variation was $0.12^\circ$C. A similar trend was observed for salinity, with mean differences of  0.13 kg/m$^3$, 0.06 kg/m$^3$, and 0.06 kg/m$^3$ when measured at the surface, 11 meters, and 24 meters below the surface, respectively, and \ct{a depth-averaged difference} 
    of 0.07 kg/m$^3$.} 
    Assuming normally distributed instrumental noise and horizontal variations noise that is independent of the instrument accuracy, the total noise is estimated 
    using the root-sum-square. 
    The estimated normalized noise level is $\eta = 0.047$  for temperature and \ct{$\eta = 0.077$} for salinity, where $\eta$ is defined in \cref{sec:num_setup}.
    These estimates are comparable to the largest noise levels in the synthetic problems presented in \Cref{sec:syn_prob}.
    Based on the results shown in \Cref{fig:diff-recovery-single-double}, the inverted diffusion coefficient from each data is expected to be reliable at locations where the data exhibit the highest sensitivity, but difficult to estimate in regions where the data have a small derivative, corresponding to regions of low sensitivity.

    \subsection{Data compatibility condition}\label{sec:data_comp}
        Although the forward problem is always well-posed, some measurement data are not attainable in the inverse problem by varying only the diffusion coefficient, subject to the physical constraint that it must be positive. As a result, some data may be incompatible with the chosen model.
    To recognize and potentially avoid this, a \emph{data compatibility condition} is introduced to ensure that the solution space of the differential equation aligns with the data. Note that this differs from the \emph{feasibility conditions}, for example, for Poisson's equation with Neumann boundary conditions, which ensure the existence of solutions to the differential equation.
    
    By integrating \Cref{eq:helm_main} and applying the divergence theorem on the diffusion term, we obtain the following compatibility condition relating boundary conditions and interior values:
    \begin{align}\label{eq:compatibility_cond}
        -\int_{\partial\Omega} e^m\nabla u \cdot \textbf{n} \, ds = 
        \into \omega d(x)\left(u_{oce} - u\right) + f(x)\,dx.
    \end{align}
    Since here the domain of interest is one-dimensional, i.e., $\Omega = [0,1]$, the boundary integral can be evaluated as $-\int_{\partial\Omega} e^m\nabla u \cdot \textbf{n} \, ds = -\left[e^m \frac{du}{dx}\right] \big|_0^1$. 
    When using a Neumann boundary condition, as is the case for the salinity model, the flux imposed at the boundaries must therefore match the integral on the right-hand side. 
    If the data are a solution to the governing equations, condition (\ref{eq:compatibility_cond}) will be satisfied when $u_d$ is used in place of $u$. 
    We note that in a one-dimensional domain, a similar condition can be obtained over any interval within the domain, including, notably, points where the solution has a zero derivative and the flux is therefore known to be zero.
    
    Because the diffusion parameter is strictly positive, the flux term inherits its sign from the derivative of $u$ at the boundary.
    Since $\nabla u$ can often be inferred directly from boundary measurements, this flux term at the boundary is effectively sign-constrained for every feasible $m$.
    If the prescribed sign of the boundary flux is inconsistent with the sign implied by the integral on the right-hand-side of \Cref{eq:compatibility_cond}, the data lie outside the range of the forward mapping and are incompatible with the model. 
    Detecting this sign mismatch can serve as a useful diagnostic for the model, potentially indicating missing dynamics in the governing equation.
    
    The data shown in \Cref{fig:real_lake_data} reach 24 meters below the surface, but the deepest location of the lake reaches 30 meters. 
    Therefore, to determine the bottom boundary condition for the salinity problem, \Cref{eq:compatibility_cond} was used in combination with the salinity flux at the top boundary mentioned in the previous subsection.
    Evaluating \Cref{eq:compatibility_cond} with $u = u_d$ gave the flux boundary condition at the bottom as approximately $1.32\times 10^{-3}$ kg/m$^2$s, which ensures that the data are compatible with the proposed model.

    \subsection{Numerical results}\label{sec:app_restuls}
    \begin{figure*}[t]
        \centering
        \begin{overpic}[width=0.49\textwidth]{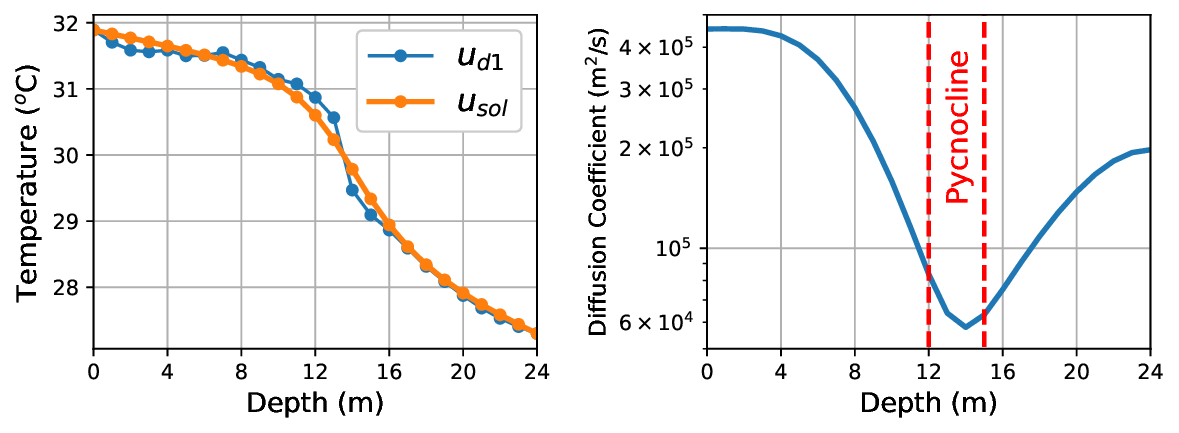}
            \put(10,9){\normalsize A.}
            \put(62,9){\normalsize B.}
        \end{overpic}
        \begin{overpic}[width=0.49\textwidth]{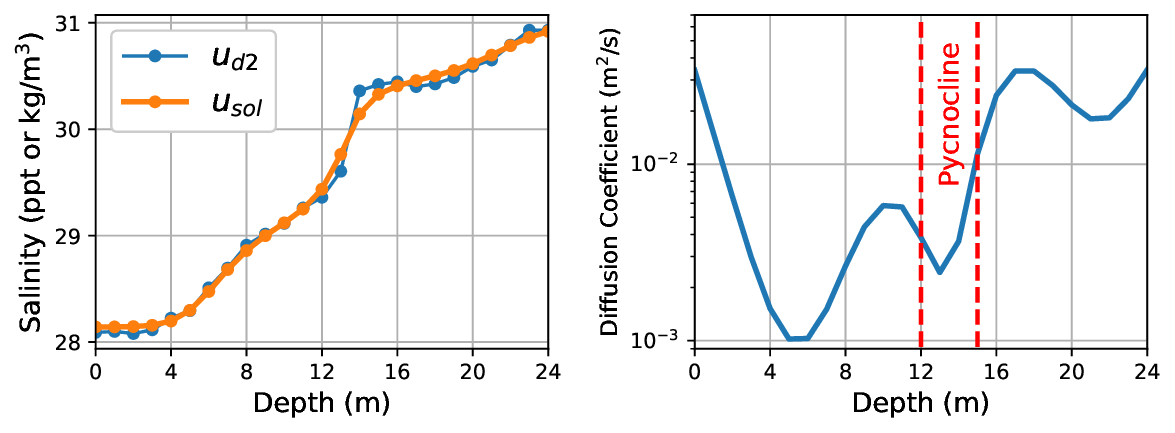}
            \put(42,9){\normalsize C.}
            \put(92.5,9){\normalsize D.}
        \end{overpic}
        \caption{
        Panels A and B: Temperature data optimization with real marine lake data. The regularization parameter obtained from the L-curve criterion is $\gamma = 7.94\times 10^{-5}$. Panel A shows the data and recovered temperature and Panel B shows the reconstructed diffusion coefficient. Panels C and D: Salinity data optimization with real marine lake data. The regularization parameter obtained from the L-curve is $\gamma = 1.58\times 10^{-5}$.  Panel C shows the data and recovered salinity and Panel D shows the reconstructed diffusion coefficient.
        }
        \label{fig:single_data_opt_real}
    \end{figure*}

    In \Cref{fig:single_data_opt_real}, we present the
    recovered solution to the forward problem and the diffusion coefficient
    obtained using either temperature or salinity data alone. Both recovered solutions provide a very good fit to the 
    data, suggesting consistency with the governing equation.
    However, the diffusion/mixing coefficients inferred from temperature and salinity are drastically different in shape and magnitude, contradicting our expectation that diffusion/mixing processes have similar effects on temperature and salinity.
    The diffusion coefficient inferred from the temperature shows the expected trend of being larger near the surface, where convective motion and wind-induced flows are largest. 
    It also dips near the pycnocline, where density stratification hinders motion, and increases again near the bottom, where the density is closer to being constant. 
    However, the inferred values of the diffusion coefficient are unexpectedly large when compared to previous estimates \cite{blanchette2020marine}, suggesting potentially missing information in the model. 
    The present adjoint-based algorithm tries to compensate by adjusting the diffusion coefficient to account for those missing contributions, leading to an overestimate of its magnitude.
    It is also useful to note that, especially near the surface, the temperature derivative is small, indicating an area of low sensitivity.
    
    The reconstructed diffusion coefficient from salinity data has a reasonable magnitude, but the 
    oscillations observed below 5m are unlikely to be physical. 
    The magnitude of the oscillation below 5m reaches the same level as that of the mixing at the surface, indicating the same level of mixing at the surface as well as near the bottom of the lake.
    These observations hint at some missing information in the model, particularly near the bottom. 
    For the salinity profile, the derivative near the very top and below the pycnocline is also small, again resulting in areas of low sensitivity.

    \begin{figure*}[h!]
        \centering
        \vspace{0.1cm}
        \begin{overpic}[width=\linewidth]{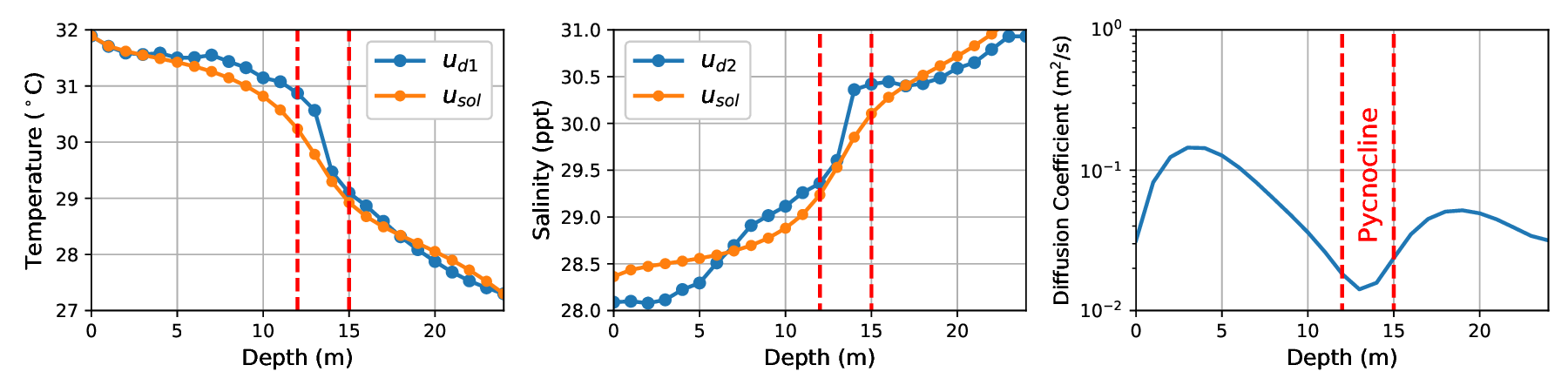}
            \put(7.6,24){\normalsize A.}
            \put(41,24){\normalsize B.}
            \put(74.3,24){\normalsize C.}
        \end{overpic}
        \caption{The recovery of both temperature (Panel A) and salinity (Panel B) when optimizing conjointly. 
        The inferred inverse solution (Panel \ct{C}) shows low diffusivity in the pycnocline, where there is a steep density gradient. 
        }
        \label{fig:double_data_opt_real}
    \end{figure*}
    
    \Cref{fig:double_data_opt_real} displays the inferred diffusion coefficient when temperature and salinity profiles are assimilated simultaneously.
    This inferred diffusion coefficient has a reasonable magnitude, of the order of $0.1$m$^2$/s, and a smoother, more physically realistic vertical structure.
    As expected on the physical grounds that mixing is greatest near the surface, the
    inferred diffusion coefficient is largest above the pycnocline.
    It then drops significantly in the pycnocline (between 12 and 15 m), where both temperature and salinity data exhibit the steepest gradient, suggesting the highest sensitivity and the most reliable parameter inversion. 
    The diffusion coefficient then increases again at greater depth, but to a lesser extent than near the surface. The forward solutions generally match the data trend well in most areas, except for the surface salinity data.
    
    Extensive tests indicate that the partial mismatch between the data and the forward solution, and the differences in inferred diffusion coefficients when using single data sets, are not numerical artifacts. 
    This dataset presents a difficult inversion problem, with a relatively high noise level and regions of low sensitivity to data.
    Despite these limitations, the present adjoint-based algorithm indicates that the screened-Poisson equation used here may benefit from additional physics, such as biological activities, a more detailed model of the ocean-lake exchanges, or other meteorological contributions. 
    Using the present algorithm to infer the diffusion coefficients provides a powerful tool in assessing model accuracy, and exploring these extensions offers a promising avenue toward a more comprehensive model.

\section{Conclusion} \label{sec:conclusion}

This work introduced a robust approach to estimate the spatially varying diffusion coefficient in stratified marine lakes using a regularized, DE-constrained optimization framework. 
This approach is also adaptable to non-stratified lakes.
The adjoint-based approach integrates a specialized governing equation \cite{blanchette2020marine}, an adjoint-based algorithm \cite{petra2011}, regularization techniques \cite{hansen2001lcurve,hilt1977ridge}, systematic sensitivity analyses, and multiple observational datasets (temperature and salinity).
Building on the well-established adjoint-based method, the novelty of the present work lies in combining the previously mentioned components in a limnological context to infer a spatially varying effective diffusion coefficient from data. In contrast, this coefficient was previously approximated only from physical intuition or simplified models \cite{blanchette2020marine,riley1988minlake,de2016estimation,botte2000numerical, filatov2011field}.
The present numerical study demonstrated that the inverse-problem formulation yields meaningful reconstructions of the diffusion coefficient from both real and synthetic data. 
Moreover, leveraging the sensitivity analysis inherent in the adjoint-based formulation (\ct{\Cref{sec:sens_analysis}}) demonstrated an avenue to assess the quality of the inferred parameter throughout the optimization process.
The results presented highlight both the strengths and limitations of this approach, notably its effectiveness in mitigating ill-posedness and achieving stable parameter inversion through multi-data optimization accompanied by a manageable increase in computational cost.

In this paper, the model problem was discretized using the FEM with first-order Lagrange polynomials as basis functions.
Numerical experiments confirmed that the method exhibits second-order accuracy with respect to spatial resolution, as expected from the choice of finite element discretization. 
Additionally, a Tikhonov regularization term was included in the cost function, with its prefactor being determined using the L-curve method. 
The regularization study presented in this paper demonstrated the effectiveness of the L-curve method in selecting an appropriate Tikhonov regularization parameter, balancing the trade-off between solution stability and data fidelity.

To assess robustness, noise at various levels was added to the data. The results showed that the inversion error, $\lVert e^m - e^{m_{true}}\rVert$, is not very
sensitive to the choice of regularization parameter, showing the robustness of the L-curve method 
even in high-noise regimes.
Error analysis shows that reconstruction quality degrades with increasing noise level, particularly in regions with low sensitivity, leading to reduced reconstruction accuracy. 
Outside regions of low sensitivity, such as where the solution gradient is high, good reconstruction was observed up to a noise level of about 0.05.
By incorporating multiple datasets with complementary sensitivity profiles, reconstruction quality improved, particularly in regions where individual datasets exhibit low sensitivity.
To quantify the benefit of using multiple datasets, the spectrum of the reduced Hessian misfit was analyzed. It exhibited larger dominant eigenvalues compared to the single-data case.
Although the larger spectrum required more optimization iterations, solving the joint problem yielded a more accurate reconstruction.

Furthermore, the present DE-constrained optimization approach was contrasted with physics-informed neural networks (PINNs).
While PINNs offer a promising alternative with implementation flexibility, they also pose challenges in hyperparameter tuning, interpretability, and training instability.
On the other hand, the present DE-constrained optimization approach achieves comparable reconstructions with fewer computational resources,  though it requires more complex mathematical derivations.
Numerical studies showed that both methods demonstrate strong capabilities in solving model discovery problems, each with trade-offs between interpretability, computational cost, and implementation complexity (at least in a small-scale regime).

Lastly, the adjoint-based method was applied to real-world data obtained in a stratified marine lake, Jellyfish Lake, in Palau. 
To ensure the existence of an inverse solution, a \emph{data compatibility condition} was derived and used as a diagnostic to ensure the measurement dataset was compatible with the differential equation model.
The results presented showed good reconstruction from the temperature and salinity data.
However, inconsistencies in the magnitude of the inferred diffusion coefficient near the lake bottom suggested that the differential equation model could benefit from further refinement.
For instance, the current model does not account for sediment porosity or interactions at the lake floor.
Incorporating additional measurements and lake morphometry data, as well as further investigation of the forcing terms for temperature and salinity, will be necessary to improve model quality and enhance reconstruction accuracy.

In future work, this method will be extended to a time-dependent version of the parameter reconstruction problem and incorporate time-dependent data. 
In theory, this is a straightforward extension \cite{ho2024inv} of the method presented here. In practice, however, the resulting linear systems are much larger, and fine-grained real data are harder to obtain and use correctly. 
As noted in the previous section, 
coarser monthly measurements are available; for preliminary results, readers are referred to \cite{ho2024inv}.
Future work will also focus on quantifying the uncertainty in the reconstruction using a Bayesian inference framework. 
In this context, it will be essential to account for uncertainties within the model stemming from multiple parameter uncertainties and model error.
In addition, one could couple the adjoint-based approach with symbolic regression \cite{pettit2025disco,petersen2019deep}, which can optimize for a physically interpretable forcing term. 
Finally, a long-term objective is to apply the adjoint-based method to more complex dynamics to assess its broader applicability across different application problems, such as different ocean regions. 





This work was partially supported by the National Science Foundation under Grant No. DMS-1840265.





\printcredits
\section*{Declaration of competing interest}
The authors declare that they have no known competing financial interests or personal relationships that could have appeared to
influence the work reported in this paper.
\section*{Acknowledgement}
We thank Gerda Ucharm, Sharon Patris, and Lori Colin at the Coral Reef Research Foundation, Palau, for generously providing data and methods from their long-term marine lake monitoring program in Palau that made this work possible.

\bibliographystyle{cas-model2-names}

\bibliography{cas-refs}



\end{document}